\documentclass[12pt,preprint]{aastex}

\newcommand{\lw}[1]{\smash{\lower2.ex\hbox{#1}}}
\def\kms{km~s$^{-1}$}

\def\Vlsr{$V_{\rm LSR}$}

\def\kms{\mbox{km~s$^{-1}$}}
\def\cmc{cm$^{-3}$}
\def\cmq{cm$^{-2}$}

\def\um{350\,$\mu$m}

\def\Msun{\mbox{$M_\odot$}}
\def\Lsun{\mbox{$L_\odot$}}

\def\Vlsr{$V_{\rm LSR}$}
\def\Vsys{$V_{\rm sys}$}

\def\nodata{$\cdot\cdot\cdot$}

\def\Snu{$S_{\nu}$}
\def\Fnu{$F_{\nu}$}

\def\Er{$E_{\rm r}$}

\def\wat{H$_2$O}
\def\amm{NH$_3$}

\def\ccsI{CC$^{34}$S}
\def\isolated{N$_2$H$^+$ (1--0) F$_1$,F$=$0,1--1,2}
\def\tisolated{F$_1$,F $=$ 0,1--1,2}
\def\tbrightest{F$_1$,F $=$ 2,3--1,2}
\def\co{$^{12}$CO}

\def\NtwoH{N$_2$H$^+$}
\def\HtCOp{H$^{13}$CO$^+$}
\def\HCeOp{HC$^{18}$O$^+$}
\def\CtHt{C$_3$H$_2$}
\def\Tbol{$T_{\rm bol}$}
\def\Lbol{$L_{\rm bol}$}

\def\meanNHmol{$\langle N({\rm H_2})\rangle$}

\def\nHmol{$n({\rm H_2})$}

\def\pbeam{beam$^{-1}$}

\def\dVint{\mbox{$\Delta v_{\rm int}$}}
\def\dVfwhm{\mbox{$\Delta v_{\rm FWHM}$}}
\def\meandVint{\mbox{$\langle\Delta v_{\rm int}\rangle$}}

\def\Cs{\mbox{$c_{\rm s}$}}
\def\Vinf{\mbox{$v_{\rm inf}$}}

\def\dVthm{\mbox{$\Delta v_{\rm thm}$}}

\def\dVnth{\mbox{$\Delta v_{\rm nth}$}}
\def\meandVnth{\mbox{$\langle\Delta v_{\rm nth}\rangle$}}

\def\Tb{\mbox{$T_{\rm B}$}}
\def\Tk{\mbox{$T_{\rm k}$}}
\def\Tr{\mbox{$T_{\rm rot}$}}
\def\Tr21{\mbox{$T_{\rm r,21}$}}
\def\Tmb{\mbox{$T_{\rm mb}$}}
\def\Tsys{\mbox{$T_{\rm sys}$}}
\def\Tmb{\mbox{$T_{\rm mb}$}}
\def\Tastar{\mbox{$T_{\rm A}^{\ast}$}}
\def\Tsb{\mbox{$T_{\rm sb}$}}
\def\Tb{\mbox{$T_{\rm b}$}}
\def\Td{\mbox{$T_{\rm d}$}}
\def\Tex{\mbox{$T_{\rm ex}$}}
\def\JTex{\mbox{$J_{\nu}(T_{\rm ex})$}}
\def\Tbg{\mbox{$T_{\rm bg}$}}
\def\JTbg{\mbox{$J_{\nu}(T_{\rm bg})$}}
\def\Fco{\mbox{$F_{\rm CO}$}}

\def\FcocLbol{\mbox{$F_{\rm CO}\cdot c/L_{\rm bol}$}}
\def\Cs{\mbox{$c_{\rm s}$}}
\def\Reff{\mbox{$R_{\rm eff}$}}
\def\Rcent{\mbox{$R_{\rm cent}$}}

\def\NNtwoH{\mbox{$N_{\rm N_2H^+}$}}
\def\NHtCOp{\mbox{$N_{\rm H^{13}CO^+}$}}
\def\Nccs{\mbox{$N_{\rm CCS}$}}

\def\NCtHt{\mbox{$N_{\rm C_3H_2}$}}
\def\tff{\mbox{$t_{\rm ff}$}}
\def\tdyn{\mbox{$t_{\rm dyn}$}}
\def\tpstar{\mbox{$t_{\rm protostar}$}}
\def\tcore{\mbox{$t_{\rm core}$}}
\def\taut{$\tau_{\rm tot}$}

\def\tauccs{$\tau_{\rm CCS}$}
\def\tauctht{$\tau_{\rm C_3H_2}$}
\def\tauhtcop{$\tau_{\rm H^{13}CO^+}$}

\def\Menv{\mbox{$M_{\rm env}$}}

\def\MLTE{\mbox{$M_{\rm LTE}$}}
\def\Mvir{\mbox{$M_{\rm vir}$}}

\def\Mcbd{\mbox{$M_{\rm cbd}$}}

\def\gf{\mbox{GF\,9-2}}
\def\sharc{\mbox{SHARC\,{\sc II}}}

\def\Omegasrc{\mbox{$\Omega_{\rm s}$}}
\def\Omegabeam{\mbox{$\Omega_{\rm b}$}}

\def\thetamaj{$\theta_{\rm maj}$}
\def\thetamin{$\theta_{\rm min}$}
\def\thetahpbw{$\theta_{\rm HPBW}$}

\def\Dlambdafull{$D^{\rm F}_{\lambda \,{\rm SD}}$}
\def\Dlambdanyquist{$D^{\rm N}_{\lambda \,{\rm SD}}$}
\def\uv{{\it u-v}}
\def\RadpSimSoltn{$\rho (r)\propto r^{-2}$}
\def\RadpInf{$\rho (r)\propto r^{-3/2}$}
\def\XNtwoH{\mbox{$(9.5\pm 5.0)\times 10^{-10}$}}
\def\XNtwoHEE{\mbox{$(9.5\pm 5.0)$E--10}}
\def\XNtwoHE{\mbox{9.5E--10}}

\def\XHtCOp{\mbox{$(8.5\pm 4.7)\times 10^{-12}$}}
\def\XHtCOpEE{\mbox{$(8.5\pm 4.7)$E--12}}

\def\Xccs{\mbox{$(5\pm 3)\times 10^{-10}$}}
\def\XccsEE{\mbox{$(5\pm 3)$E--10}}

\def\Xamm{\mbox{$(2\pm 1)\times 10^{-8}$}}
\def\XammEE{\mbox{$(2\pm 1)$E--8}}

\def\XCtHt{\mbox{$(2\pm 1)\times 10^{-10}$}}
\def\XCtHtEE{\mbox{$(2\pm 1)$E--10}}

\def\lesssim{\mathrel{\hbox{\rlap{\hbox{\lower4pt\hbox{$\sim$}}}\hbox{$<$}}}}
\def\gtrsim{\mathrel{\hbox{\rlap{\hbox{\lower4pt\hbox{$\sim$}}}\hbox{$>$}}}}

\slugcomment{Accepted by ApJ on 2006 August 13}

\shorttitle{Initial Conditions for Gravitational Collapse of a Core}
\shortauthors{Furuya et al.}

\begin{document}


\title{The Initial Conditions for Gravitational Collapse of a Core:
An Extremely Young Low-Mass Class 0 Protostar GF\,9-2}


\author{Ray S. FURUYA\altaffilmark{1}}
\affil{Division of Physics, Mathematics, and Astronomy, California Institute of Technology}
\email{rsf@subaru.naoj.org}

\author{Yoshimi KITAMURA\altaffilmark{2}}
\affil{Institute of Space and Astronautical Science, Japan Aerospace Exploration Agency}
\email{kitamura@isas.jaxa.jp}

\and

\author{Hiroko SHINNAGA\altaffilmark{3}}
\affil{Caltech Submillimeter Observatory, California Institute of Technology}
\email{shinnaga@submm.caltech.edu}


\altaffiltext{1}{Present address: Subaru Telecope, National Astronomical 
Observatory of Japan, 650 North A'ohoku Place, Hilo, HI 96720}
\altaffiltext{2}{3-1-1 Yoshinodai, Sagamihara, Kanagawa 229-8510, Japan}
\altaffiltext{3}{111 Nowelo Street, Hilo, HI 96720}
\altaffiltext{$\ast$}{\bf\large The preprint including figures with the
original quality is available at 
\texttt{http://subarutelescope.org/staff/rsf/publication.html}}


\begin{abstract}
We present a study of the natal core harboring the class 0 protostar \gf\ 
(bolometric luminosity $\simeq 0.3$ \Lsun, 
bolometric temperature $\lesssim 20$ K; Wiesmeyer et al.) 
in the filamentary dark cloud GF\,9 (distance $=$ 200 pc) using the
Nobeyama 45\,m, CSO 10.4\,m telescopes, and the OVRO mm-array.
\gf\ stands unique in the sense that it shows \wat\ maser emission (Furuya et al.), 
a clear signpost of protostar formation, 
whereas it does not have a high-velocity large-scale molecular outflow
evidenced by our deep search for \co\ wing emission.
These facts indicate that \gf\ core is early enough after star formation 
so that it still
retains some information of initial conditions for collapse.
Our \um\ dust continuum emission image revealed the presence of a 
protostellar envelope 
with an extent of $\simeq$ 5400 AU in the center of 
a molecular core $\simeq$ 0.08 pc in size.
The mass of the envelope is $\simeq 0.6$ \Msun\ from the \um\ flux density,
while LTE mass of the core is $\simeq 3$ \Msun\ 
from \NtwoH, \HtCOp, CCS, and \amm\ line observations.
Combining visibility data from the OVRO mm-array and the 45\,m telescope, 
we found that the core has a radial density profile of
$\rho(r)\propto r^{-2}$ for $0.003\lesssim r/{\rm pc}\lesssim 0.08$ region.
Molecular line data analysis revealed that 
the velocity width of the core gas increases inward, 
while the outermost region maintains a velocity dispersion of a few times 
of the ambient sound speed.
The broadened velocity width can be interpreted as infall.
Thus, the collapse in \gf\ is likely 
to be described by an extension of 
the Larson-Penston solution for the
period after formation of a central star.
We derived the current mass accretion rate of $\simeq 3\times 10^{-5}$ 
\Msun\ year$^{-1}$ from infall velocity of $\simeq$ 0.3 \kms\ at $r\simeq$ 7000 AU.
Furthermore, we found evidence that a protobinary is being formed 
at the core center. 
All results suggest that \gf\ core has been undergoing gravitational collapse
for $\lesssim$ 5000 years since the formation of central protostar(s), and that 
the unstable state initiated the collapse $\simeq 2\times 10^5$ years
(the free-fall time) ago. 
\end{abstract}





\section{Introduction}

Stars form inside dense ($\simeq 10^5$ \cmc) and compact ($\simeq 0.1$ pc) molecular cores
(e.g., Myers \& Benson 1983; Mizuno et al. 1994).
One of the roles of such cores is to provide material for newly forming stars
through dynamical accretion caused by gravitational collapse of the cores.
For isolated low-mass star formation, 
two extreme paradigms for collapse of a core have been proposed.
One is a slow, quasi-static model, while the other extreme
is the dynamic, turbulent one.
The former is represented by the Shu's similarity solution (Shu 1977), 
and the latter by the Larson-Penston's solution (Larson 1969; Penston 1969)
which was extended for the period after a protostar formation
by Hunter (1977) and Whitworth \& Summers (1985). 
Hereafter, we refer to the latter extended solution as the Larson-Penston's solution.
The two solutions are discriminated by radial density and velocity profiles of 
the core at $t =$ 0.  
The Shu's solution (1977) sets the density profile of 
$\rho(r)=\frac{c_{\rm s}^2}{2\pi G}~r^{-2}$ and is static ($v(r) =0$) at $t=0$ 
where $c_{\rm s}$ and $G$ are the isothermal sound speed and the 
gravitational constant, respectively.  
At $t = 0$, the collapse of the core begins at the center, 
which produces an expansion or rarefaction wave that propagates outward 
with the local sound speed. 
This model is referred to as the ``inside-out'' collapse 
(Shu 1977; Shu, Adams \& Lizano 1987).  
On the other hand, the Larson-Penston's solution approaches the profile of 
$\rho(r)\simeq 4.4~ \frac{c_{\rm s}^2}{2\pi G}~r^{-2}$ and in infalling state 
($v(r) \approx -3.28~c_{\rm s}$) when a protostar is about to be born 
($t\rightarrow -0$, Hunter 1977; Shu 1977; Foster \& Chevalier 1993).
Thus the Larson-Penston solution, referred to as the ``runaway'' collapse, 
predicts a 48 times higher mass infall rate than the Shu's. 
At $t >$ 0, the both solutions have a free-fall region of \RadpInf\
which expands with sound speed for the Shu solution and 
with supersonic velocity for the Larson-Penston that
accompanies no rarefaction wave.\par

Molecular line and dust continuum observations have identified a number of
gravitationally bound starless
cores (e.g., Mizuno et al. 1994), referred to as ``preprotostellar'' cores
(Ward-Thompson et al. 1994).
Thermal dust continuum emission imaging at mm and submm wavelengths 
revealed that radial density profiles of the preprotostellar cores
have much flatter profiles than $\rho (r)\propto r^{-2}$ 
at radii smaller than a few thousand AU 
(e.g., Ward-Thompson et al. 1994; Andr\'e et al. 1996).
Molecular line studies measuring velocity field of the cores
are also powerful tool to verify collapse models (e.g., Zhou 1992).
For this purpose, 
it is the most important to search for a star-forming core which retains
the initial conditions for the gravitational collapse, 
namely, a star-forming core at an extremely early evolutionary
phase before dispersal by a large-scale molecular outflow.
Furuya et al (1999, 2000) proposed that the intermediate-mass
class 0 source S106\,FIR is such an ideal core, 
because the source is characterized by 
(1) a cold SED (Richer et al. 1993), (2) \wat\ maser emission, i.e., 
strong evidence for the presence of a protostar, and 
(3) absence of a large-scale high-velocity CO outflow.\par

We have successfully identified a low-mass counterpart, \gf\ 
by performing multi-epoch \wat\ maser survey
(Furuya et al. 2001, 2003) and \co\ outflow mapping survey 
toward the selected low-mass protostars with the Nobeyama 45\,m, SEST 15\,m, 
and CSO 10.4\,m telescopes (R. S. Furuya et al. in preparation).
In our sample, \gf\ is the only protostar satisfying the above characteristics of (1)--(3).
This protostar is located in the east-west filamentary dark cloud GF\,9 
where several YSOs are located every $\simeq 0.75$ pc in the plane of the sky
(Schneider \& Elmegreen 1979; Wiesemeyer 1997; 
the filament is cross-identified as Barnard\,150 and Lynds\,1082.
The distance to \gf\ is estimated to be 200 pc by Wiesemeyer (1997).
Wiesemeyer et al. (1997, 1999) identified \gf\ as class 0 and obtained 
bolometric temperature (\Tbol) of $\leq$ 20 K and 
bolometric luminosity (\Lbol) of $0.3 L_{\odot}$.
We believe that \gf\ offers
unique opportunity to investigate the initial conditions 
for gravitational collapse of a star-forming core. 
This paper presents a detailed observational study of \gf. 
In particular,
we combined mm-interferometer and single-dish telescope 
visibility data to obtain high fidelity images of the core.

\section{Observations}

Toward GF\,9-2, we observed dust continuum emission with the Caltech Submillimeter Observatory
\footnote{The Caltech Submillimeter Observatory is operated by
the California Institute of Technology under funding from
the US National Science Foundation (AST 02-29008).}
(CSO) 10.4\,m telescope ($\S\ref{ss:csoobs}$) and molecular line emission with the
Nobeyama Radio Observatory\footnote{Nobeyama Radio Observatory
is a branch of the National Astronomical Observatory,
operated by the Ministry of Education, Culture, Sports, Science and Technology, Japan.}
(NRO) 45\,m telescope ($\S\ref{ss:NROobs}$).
We also carried out a deep search for molecular outflow using 
the 10.4\,m and 45\,m telescopes and
used cm radio continuum data from 
the NRAO\footnote{The National Radio Astronomy Observatory is a facility of the National Science Foundation 
operated under cooperative agreement by Associated Universities, Inc.} Very Large Array (VLA) 
archive data base to search for thermal radio jet ($\S\ref{ss:vlaobs}$).
The parameters for single-dish and interferometer observations are 
summarized in Tables \ref{tbl:sdobs} and \ref{tbl:cont_obs}, respectively.\par

\subsection{Caltech Submillimeter Observatory 10.4\,m Telescope} 
\label{ss:csoobs}

\subsubsection{Mapping of \co\ (3--2) Line Emission} 
\label{ss:csolineobs}

The \co\ (3--2) line observations were done using 
the SIS receiver with position-switching mode in 2002 December.
We used the 50 MHz bandwidth acousto-optical spectrometer (AOS).
During the observations, the zenith opacities at 225 GHz ($\tau_{\rm 225~GHz}$) 
ranged from 0.037 to 0.12.
The telescope pointing was checked by observing the \co\ (3--2) emission toward CRL 2688
every 1.5 hours, and the accuracy was better than $5\arcsec$.
The observing grid was centered on the position of the 3\,mm continuum source 
reported in Wiesemeyer et al. (1999)
(R.A.$_{\rm J2000}=20^{\rm h}51^{\rm m}29\farcs 5$,
Dec.$_{\rm J2000}=60\degr 18\arcmin38\farcs0)$.
In the data reduction, we removed spectra
obtained with elevation angle of $\lesssim 30\degr$ or
with \Tsys\ $\geq$ 800 K in single sideband (SSB).\par

\subsubsection{Imaging of \um\ Continuum Emission with \sharc} 
\label{ss:sharc2obs}

The \sharc\ camera (Dowell et al. 2003) was used to
obtain a 350 $\mu$m continuum emission map in the box and Lissajous scan modes 
(Kov\'acs 2006) in 2004 June.
We used the Dish Surface Optimization System (DSOS) which actively corrects
gravitational deformation of the main reflector of the telescope (Leong 2005).
During our observations, $\tau_{\rm 225~GHz}$ ranged from 0.045 to 0.08.
The telescope focusing and pointing were checked every an hour by
observing Neptune and Uranus.
The measured beam size in HPBW was
$(8\farcs 9\pm 0\farcs 2)\times (8\farcs 0\pm 0\farcs 4)$ at P.A.$=$ 93\degr.
We estimated the positional accuracy of $\lesssim 4\arcsec$
from maps of the pointing sources.
For flux calibrations, we used the pointing data of Neptune
that satisfied the condition of $\tau_{225\rm GHz}\times {\rm airmass}\le 0.1$.
We obtained a mean peak voltage ($V_{\rm peak}$) of
$8.25\pm 1.5$ $\mu$Volt toward Neptune over a pixel aperture 
($A_{\rm pixel}=4\farcs9\times 4\farcs8$) of the detector.
The measured voltage was converted into source intensity as
$\frac{V_{\rm peak}}{A_{\rm pixel}}\propto\frac{S_{\nu}}{\Omega_{\rm b}}$
where $S_{\nu}$ denotes flux density per beam and \Omegabeam\ beam solid angle. 
Since the brightness temperature of $63$ K and the angular diameter of $2.6\arcsec$ 
for Neptune led to $S_{\nu}$ of 98.0 Jy at \um, 
we obtained a conversion factor of 3.2$\times 10^{-6}$ Jy \pbeam\ Volt$^{-1}$. \par

\subsection{Nobeyama 45\,m Telescope} 
\label{ss:NROobs}

\subsubsection{Observations of \co\ (1--0), \HtCOp\ and \HCeOp\ (1--0), and \CtHt\ $2_{12}-1_{01}$ Lines}
\label{ss:ssb_obs}

We carried out simultaneous observations of \co\ (1--0) and \HtCOp\ (1--0) lines
using two SIS receivers (S100 and S80) with SSB filters
in dual polarization mode in 2000 April.
For the backend, we used the 8 arrays of AOS.
All of the 45\,m telescope observations 
were made in position switching mode,  
using an off-position of 
($\Delta\alpha$, $\Delta\delta)=(-8\farcm0,+10\farcm0)$ to the 3\,mm source.
The pointing accuracy was better than $3\arcsec$.
The daily variation of the receiver gains was estimated to be 
less than 10\% in peak-to-peak, 
leading the corrected absolute flux uncertainty of $15\%$. In 2005 April, 
we made the simultaneous observations of the \HtCOp\ and \HCeOp\ (1--0) 
lines to measure optical depth of the \HtCOp\ line toward the 3\,mm continuum source.
At the same time, we also observed the ortho-\CtHt\ $2_{12}-1_{01}$ line.

\subsubsection{Mapping Observations of \co\ (1--0),  \HtCOp\ (1--0) and SiO (2--1) Lines with BEARS}
\label{ss:BEARSobs}

The mapping observations of the $^{12}$CO (1--0), \HtCOp\ (1--0) and 
SiO (2--1) $v=0$ lines 
were carried out with the 25 BEam Array Receiver System (BEARS) 
in 2001 January and April.
For backend, we used digital spectrometers. 
We adopted correction factors provided by the observatory for 
relative gain differences among the 25 beams.
The \co\ and \HtCOp\ line data taken with the S100/S80 receivers
with the SSB filters ($\S\ref{ss:ssb_obs}$)
were used to determine the antenna temperature (\Tastar) 
scale for the BEARS which is not equipped with SSB filters.
The telescope pointing accuracy was better than $3\arcsec$,
and the uncertainty of temperature scale calibration was better than $20\%$.

\subsubsection{Observations of \NtwoH\ (1--0), CCS $4_3-3_2$ and CC$^{34}$S $4_3-3_2$ Lines}

In 2003 December, we made the simultaneous \NtwoH\ (1--0) and CCS $4_3-3_2$ line observations
with a grid spacing of $40\arcsec$, full-beam sampling for the CCS line.
Subsequently, we carried out deeper integration of the \NtwoH\ line 
with a spacing of 20\arcsec, a full-beam sampling for the line.
In addition, we observed CC$^{34}$S 4$_3-3_2$ line toward the 3\,mm continuum
source to estimate optical depth of the CCS line.
The pointing accuracy and the uncertainty of flux calibration were
the same as in $\S\ref{ss:BEARSobs}$.

\subsubsection{Observations of \amm\ Inversion Lines at 23 GHz}

We carried out mapping observations of the four inversion lines 
of \amm\ molecules at 23--24 GHz using the HEMT
receiver (H20) with a dual circular polarization mode in 2001 April.
We used the eight AOS arrays to receive both the polarizations 
of the 4 transitions and added the dual polarization data
for each transition to improve S/N ratio.
The telescope pointing accuracy was better than $5\arcsec$, 
and the uncertainty of temperature calibration was better than $20\%$.

\subsection{Owens Valley Radio Observatory Millimeter Array} 
\label{ss:ovroobs}

The aperture synthesis observations at 3\,mm were carried out 
using the six-element OVRO mm array in the period 
from 2003 September to 2004 May with the 4 array configurations;
3 full-tracks at the C configuration, 5 at L, 2 at H, 
and 2 partial tracks at E (Table \ref{tbl:cont_obs}).
The resultant projected baseline length ranged from 12.8 m to 224 m;
our OVRO observations are insensitive to structures extended more than $54\arcsec$,
corresponding to 0.052 pc at d $=$ 200 pc.
We used the 3\,mm continuum source position reported in Wiesemeyer et al. (1999) 
as the phase tracking center.
We observed the N$_2$H$^+$ (1--0) line in the upper sideband and the
H$^{13}$CO$^+$ (1--0) line in the lower sideband.
The velocity resolutions were 0.402 and 0.108 \kms\ for the \NtwoH\ and
\HtCOp\ lines, respectively.\par

We used 3C\,273, 3C\,454.3 and 3C\,84 as passband calibrators and J2038$+$513
as a phase and gain calibrator.
The flux densities of J2038$+$513 were determined from observations of Uranus and Neptune.
The measured flux densities were fairly stable in the range of
1.37 -- 1.43~Jy at 86.7 GHz and of 1.26 -- 1.38~Jy at 93.2 GHz.
The uncertainty in our flux calibration was estimated to be 10\%. 
To minimize atmospheric phase fluctuation errors,
we removed all the \gf\ data taken in the switching cycles 
when the coherency of the calibrator was less than 70\% . 
The data were calibrated and edited using the MMA and MIRIAD packages.
For the continuum data, we merged the visibilities in both the sidebands
to construct an image; the final center frequency is 89.964 GHz.
Images of both the line and continuum were constructed 
using the AIPS package.\par

Since an interferometer does not have sensitivity to structures extended more than 
the largest fringe-spacing given by the shortest 
projected baseline length,
we used the 45\,m telescope maps to 
recover filtered-out emission in our OVRO mm-array observations
(Table \ref{tbl:combine}; Appendix \ref{as:combine}).
After combining interferometer and single-dish visibilities, 
we CLEANed maps using the task IMAGR in the AIPS package.
The parameters of the final maps are given in Table \ref{tbl:combine}.\par

\subsection{Very Large Array} 
\label{ss:vlaobs}

We used the 8435 MHz continuum emission image of \gf\ 
from the VLA archive database.
The data were originally taken by Wiesemeyer et al. in 1997 July with the
C-array configuration (Table \ref{tbl:cont_obs}).
The flux densities of calibrators of J2022$+$616 and J2055$+$613
were determined from observations of the standard calibration sources 
of 3\,C286 and 3\,C48, and the uncertainty in the final flux calibration 
was better than 20\%.

\section{Results}

\subsection{Continuum Emission}
\label{ss:rescont}

\subsubsection{Non-Detection of Free-Free Emission at 8.4 GHz}
\label{ss:rescont_ff}

No radio continuum emission from ionized gas is detected toward \gf\ through our 
VLA C-configuration search at 8.4 GHz ($\lambda=3.6$ cm). 
The attained 3$\sigma$ sensitivity is 0.11 mJy \pbeam,
leading to the upper limit of the 3.6\,cm continuum luminosity 
of 0.005 mJy kpc$^2$ within the beam. 
This upper limit is lower than the luminosity of the least luminous 
thermal cm continuum emission (see Figure 3 of Cabrit \& Bertout 1992; Furuya et al. 2003).
Since cm continuum emission toward protostars 
is usually attributed to free-free emission from
ionized gas produced by strong shocks of protostellar jets,
the lack of the cm emission in \gf\ suggests 
that a jet, if exists, is not so powerful to ionize ambient gas.\par

\subsubsection{Thermal Dust Emission at 3\,mm and \um}
\label{ss:rescont_dust}

In Figure \ref{fig:contmaps}, 
we present a comparison between the continuum emission maps at
\um\ taken with the \sharc\ camera and at 3\,mm with the OVRO mm-array.
The maps show that a 3\,mm point source is embedded in the extended \um\ emission.
The peak position of the 3\,mm emission is
R.A.$=20^{\rm h} 51^{\rm m}29.^{\rm s}827$,
Decl.$=60\degr ~18\arcmin ~38\farcs06$ in J2000.
The \um\ emission has a size of 33\farcs 0 $\times$ 22\farcs5,
corresponding to 6600 $\times$4500 AU which is
comparable to the typical size of a protostellar envelope
($\leq 6000$ AU; Mundy, Looney \& Welch 2000).
It is most likely that the \um\ emission traces a protostellar envelope, 
whereas the 3\,mm emission is more compact, perhaps a protoplanetary disk.
We assume that a protostar is located at the peak of the 3\,mm emission.
By integrating the emission inside the 3$\sigma$ contours,
we obtain the total flux densities \Snu\ (Table \ref{tbl:ResCont}).
Here we consider that the 3\,mm emission has negligible contribution from
optically thin free-fee emission because our non-detection of the 8.4 GHz 
continuum emission gives a $3\sigma$ upper limit of 0.090 mJy at 3\,mm 
for the free-free emission by extrapolating with $I_{\nu}\propto \nu^{-0.1}$.\par

Assuming that the dust continuum emission is optically thin, 
we estimate the envelope mass (\Menv) from the \um\ total flux density.
We used an equation of
$M_{\rm env}=\frac{S_{\nu} d^2}{\kappa_{\nu} B_{\nu}\left(T_{\rm d}\right)}$ where
$\kappa_{\nu}$ is the dust mass opacity coefficient, 
$T_{\rm d}$ the dust temperature, and
$B_{\nu}(T_{\rm d})$ the Planck function.
Assuming that dust and gas are well coupled, 
we adopt \Td\ of 9.5 K, a mean excitation temperature (\Tex) obtained
from the \NtwoH\ (1--0) data (described in $\S\ref{ss:col_n2h}$).
The value of $\kappa_{\nu}$ at \um\ is estimated by the usual form of 
$\kappa_{\nu}=\kappa_{\nu 0}\left(\nu/\nu_0\right)^\beta$.
Here we used two sets of the representative $\kappa_0$ and $\beta$ values 
reported in preprotostellar cores (case I) and in envelopes around class 0/I 
protostars (case II), because $\kappa_0$ is known to change with evolutionary class.
Recall that the source is very likely to be in an evolutionary phase 
between a preprotostellar core and a class 0 protostar.
Given $\kappa_0$ of 0.005 cm$^2$ g$^{-1}$ at $\nu_0=231$ GHz 
(Preibish et al. 1993, Andr\'e, Ward-Thompson, \& Motte 1996) and $\beta$ of 2.0 
(Beckwith, Henning \& Nakagawa 2000, and references therein) 
for case I, we have $\kappa$ of 0.069 cm$^2$ g$^{-1}$ at 850 GHz.
On the other hand, 
given $\kappa_0$ of 0.01 cm$^2$ g$^{-1}$ at 231 GHz
(Ossenkopf \& Henning 1994) and $\beta$ of 1.0 (Beckwith et al. 2000) for case II, 
we have $\kappa$ of 0.037 cm$^2$ g$^{-1}$ at 850 GHz.
Finally, we obtain the total masses of 
0.4 \Msun\ for case I and
0.8 \Msun\ for case II.
The inferred mass range is comparable to the least massive envelope masses 
in class 0 sources known 
to date, i.e., 
$M_{\rm env}\simeq 0.5$ \Msun\ 
for HH25\,MMS-MM and L1157\,MM 
(Andr\'e, Ward-Thompson \& Barsony 1993, hereafter AWB93).
Using the source size 
($\Omegasrc$) of 957 arcsec$^2$ inside the $3\sigma$ level contour, and \Menv,
a mean column density of H$_2$ molecules, $N$(H$_2$)$_{8\arcsec}$, is estimated to be
$4.2\times 10^{22}$ and $7.8\times 10^{22}$ \cmq\ 
for cases I and II, respectively.
These are converted into a mean optical depth of 
$\simeq 7\times 10^{-3}$ for both the cases, 
justifying our assumption of optically thin emission.

\subsection{Molecular Line Emission} 
\label{ss:rescore}

\subsubsection{Spectral Profile of the Molecular Line Emission toward the Center}
\label{ss:resspectra}

In Figure \ref{fig:spectra}, we present the spectra of 
\co\ (1--0), (3--2), \isolated, \HtCOp, \HCeOp\ (1--0), \CtHt\ $2_{12}-1_{01}$,
CCS, \ccsI\ $4_3-3_2$, and SiO (2--1) $v=0$  
lines toward the 3\,mm continuum source.
Figure \ref{fig:NtwoHspectra} represents the \NtwoH\ (1--0) spectrum
to show all the hyperfine components.
We summarize the parameters of these spectra in Table \ref{tbl:ResultsLine}.
Intrinsic velocity width (\dVint) in the table indicates the velocity width corrected 
for line broadening (where appropriate) due to
instrumental velocity resolution and optical depth of the line (Phillips et al. 1979). 
The \co\ (1--0) spectrum shows a broad line width of $\simeq 2$ \kms\ in 
full-width at half maximum (FWHM).
The \co\ (3--2) spectrum displays a blueshifted wing emission up to
\Vlsr\ $\simeq$ $-6$ \kms\ and a less prominent red wing up to $\simeq$ $+0.5$ \kms.
The spectrum has a central dip at \Vlsr\ $= -2.8$ \kms,  
while the high density gas tracers of \NtwoH, \HtCOp, \CtHt, and CCS show
their peaks around the dip of the \co\ (3--2) spectrum.
Such a coincidence indicates that the two peaks in the \co\ (3--2) 
spectrum arise from self-absorption, and not from two velocity components
along the line-of-sight.
At \co\ (1--0), the self-absorption is broad;
the peak in the redshifted side is almost missing, 
and only the blueshifted peak is seen.
Here, we define systemic velocity ($V_{\rm sys}$) of the cloud core
as $-2.5$ \kms, the peak velocity of the CCS $4_3-3_2$ line, because 
the line has a single sharp peak and is optically thin 
(described in $\S\ref{ss:col_ccs}$).
In addition, the \HtCOp\ and \NtwoH\ lines show similar line profiles
in the sense that they have stronger blue peaks,  
a typical signature of infall (e.g., Walker et al. 1986).
The \CtHt\ spectrum shows a flat-top shape.
Lastly, we point out that no molecular line emission is seen
around \Vlsr\ $= 5.7$ \kms\ where we detected
the weak \wat\ masers (Furuya et al. 2003).
This may suggest that the masers are excited in shocked region impacted
by a possible protostellar jet
because such jet-origin water masers in low-mass YSOs 
usually do not appear around their 
cloud velocities (Terebey, Vogel \& Myers 1992; Chernin 1995;
Claussen et al. 1998; Moscadelli et al. 2006).

\subsubsection{Single-Dish Maps of High Density Gas} 
\label{ss:resoverview}

Figure \ref{fig:tiimaps} illustrates total integrated intensity maps 
of the high density gas tracers of \HtCOp\ (1--0), CCS $4_3-3_2$, 
\NtwoH\ (1--0), and \amm\ lines
taken with the 45\,m telescope.
We integrated all of the hyperfine components for the nitrogen-bearing 
molecules having hyperfine structures.
The four lines show similar overall structures.
It is very likely that all of the molecular lines trace a natal cloud core of \gf.
The highest contour of the \HtCOp\ and \NtwoH\ emission show east-west (EW) elongation.
In addition, the \NtwoH\ map shows a higher degree of central 
concentration than the \HtCOp\ and CCS maps with comparable beam sizes.
The 3\,mm dust continuum emission is located at the densest innermost 
region of the core, although its position is slightly displaced to the east
of the \HtCOp\ and \NtwoH\ peaks.\par

Figure \ref{fig:NROchmaps} shows velocity channel maps of the \HtCOp\ (1--0) and 
\isolated\ lines observed with the 45\,m telescope; 
the latter transition is not blended with the other 6 hyperfine components.
The \HtCOp\ channel maps exhibit a velocity gradient
along the major axis of the core; 
the redshifted gas dominantly lies to the NE of the 3\,mm source, whereas
the blueshifted one to the SW.

\subsubsection{Combined OVRO mm array and Nobeyama 45\,m Telescope Data} 
\label{ss:combine}

Figure \ref{fig:hikaku} shows the combined \NtwoH\ and \HtCOp\ images together
with the original 45\,m and OVRO maps, and 
Figure \ref{fig:cmbchmaps} shows channel maps of the combined data.
We note that the OVRO and combined images are insensitive to 
the emission outside the fields of view of the OVRO array 
due to primary beam attenuation of the element antennas, 
and that the combined \HtCOp\ and \NtwoH\ images are
$\simeq 4$ and 15 times less sensitive compared to the 45\,m maps, respectively.
The OVRO images in Figure \ref{fig:hikaku} trace the innermost region of the core
and clearly show that both the \NtwoH\ and \HtCOp\ peaks are shifted 
to the west of the 3\,mm source. 
The combined images in Figure \ref{fig:hikaku} show that there 
is a centrally peaked feature which is elongated along EW in the smaller scale
and that intensity of the two lines decreases with increasing the radial distance 
from the center. However, contrary to Figure \ref{fig:NROchmaps}, 
one cannot see a well-defined velocity gradient in the combined channel 
maps of the \HtCOp\ emission (Figure \ref{fig:cmbchmaps}).

\subsection{Search for Molecular Outflows and Jets}
\label{ss:res_moloutflows}

\subsubsection{Search for CO Outflow}
\label{ss:res_outflows}

Figure \ref{fig:COBluRed} compares the blue- and redshifted \co\ (1--0) and (3--2) 
emission integrated over the velocity ranges shown in Figure \ref{fig:spectra}.
In the figure, no highly-collimated high-velocity ``jet-like'' outflow typically seen 
in class 0 sources (Bachiller 1996, and references therein) is identified toward the 3\,mm source, 
although formation of protostar in \gf\ core is strongly suggested by the \wat\ masers.
This is because the terminal velocity of the blue- and redshifted emission,
shifted by only $\simeq 3$ \kms\ to \Vsys\ (Figure \ref{fig:spectra}),
is too small to be interpreted as the line-of-sight components of CO outflow
velocity ($V_{\rm flow}$) in class 0 sources
(e.g., $V_{\rm flow} \simeq$ 50 \kms\ for VLA\,1623; Andr\'e et al. 1990).
Instead, a condensation of the low-velocity redshifted gas 
lies to the south of the 3\,mm source in the (3--2) map. 
A similar condensation is marginally seen in the redshifted (1--0) map, 
but seems to be a part of the large-scale natal filament. 
These red condensations might represent a compact low-velocity outflow which 
has just launched from a protostar.
On the other hand, a patchy distribution of the blueshifted (3--2) gas is observed,
partly because of the quality of spectral baselines.
The blueshifted (1--0) map shows a ridge to the NE of the 3\,mm source,  and
is connected to the EW natal filament, whereas the blueshifted (3--2) emission 
seems to be resolved into two peaks in the core center.\par

First, we estimate an upper limit to the mass of the high-velocity outflow gas. 
Since the CO emission is very likely to be optically thick, 
we consider that the peak brightness temperature gives a good
estimate of \Tex, assuming that the emission fills the beam.
We derive \Tex\ of 5.9 K from the (1--0) and 6.9 K from (3--2) transitions, 
considering the cosmic background temperature (\Tbg\ $=2.7$ K).
We attained $3\sigma$ upper limits of 12 and 79 mK for the (1--0) and (3--2)
transitions, respectively, with a velocity width of 50 \kms\
which is a typical velocity of high-velocity jet-like 
outflows associated with class 0 sources (e.g., Bachiller 1996), 
although Figure \ref{fig:spectra} shows terminal velocity is $\simeq$ 3 \kms.
Given the mean value of \Tex $=$ 6.4 K and [CO]/[H$_2$] $\simeq$ 10$^4$ 
(Dickman 1978), we convert these upper limits into 
mass per lobe ($M_{\rm lobe}$) of 
$6.7\times 10^{-5}$ \Msun\ for (1--0) and
$1.7\times 10^{-3}$ \Msun\ for (3--2).
Clearly, these 3$\sigma$ detection limits in mass are sufficient to identify any of the
CO outflows associated with class 0 sources known to date
(e.g., Richer et al. 2000).\par

Next, we estimate mass of the possible low-velocity ($\simeq$ 3 \kms) 
outflow suggested by the
weak wing emission in Figure \ref{fig:spectra} and 
by the redshifted condensation in Figure \ref{fig:COBluRed}.
The extents of the lobes are approximately
0.053 and 0.032 pc for the blue- and red ones from the 3\,mm source, respectively.
Given \Tex $=$ 6.4 K, we obtain
$M_{\rm lobe}$ of 
$1.6\times 10^{-4}$ \Msun\ for the blue and
$2.6\times 10^{-4}$ \Msun\ for the red.
These lobe masses are considerably smaller than those in class 0 sources.

\subsubsection{Non-Detection of Shock Excited SiO Emission}
\label{ss:res_sio}

Thermal SiO emission is believed to be one of the most reliable 
shock tracers associated with
molecular outflow (e.g., Codella, Bachiller \& Reipurth 1999) because
the destruction of dust grains releases silicon into the gas-phase
and silicon is rapidly oxydized to produce SiO (e.g., Schilke et al. 1997).
No evidence of outflow shocked regions is found in the SiO emission 
over a 0.22 pc $\times$ 0.14 pc region centered on the 3\,mm continuum source.
The searched area is sufficiently large to detect
such SiO emission enhanced at leading edge of the possible outflow lobe 
younger than dynamical age of 
$\lesssim 0.053$ pc/3 \kms $\simeq 1.7\times 10^4$ years.
This result suggests that the compact low-velocity outflow 
($\S\ref{ss:res_outflows}$)
is not powerful enough to excite thermal SiO emission.\par

\section{Analysis}

\subsection{Column Density of the Core} 
\label{ss:rescolumn}

\subsubsection{\NtwoH\ (1--0)} 
\label{ss:col_n2h}

To estimate the column density of \NtwoH\ (\NNtwoH),
we analyze the hyperfine structure (HFS) due to the electric 
quadrupole moment of the N nucleus. 
We performed least-squares fitting to each spectrum by applying a model 
spectrum having seven Gaussian components assumed to have the intrinsic intensity 
ratios in local thermodynamic equilibrium (LTE) 
at the limit of $\tau =0$ (Cazzoli et al. 1985).
The details of the HFS analysis and the column density calculations are
described in Appendix \ref{ass:NtwoH}.
The free parameters are \Tex, 
velocity width ($\Delta v_{\rm FWHM}$), 
LSR velocity of the brightest hyperfine (HF) component
($J=$1--0 \tbrightest\ transition at 93173.809 MHz),
and the sum of the optical depths of the components (\taut).
We have presented an example \NtwoH\ spectrum toward the 3\,mm peak position 
in Figure \ref{fig:NtwoHspectra}. We used only the single dish 
data because the velocity resolution is 
3.4 times higher than that of the combined data.
The parameters of the best-fit model profiles for the HF spectrum, 
shown by the green histogram, are summarized in Table \ref{tbl:ResultsLine}.\par

To perform HFS analysis on the combined data, 
we create a spectrum at each map pixel from the 3D cube data.
The pixel size is set to be $0\farcs9$, 
the maximum size to avoid aliasing in the Fast Fourier Transformation (FFT) 
of the combined visibilities, 
whereas that for the 45\,m data is set to be 5\arcsec.
We obtained reasonable fitting results from 262 and 1776 spectra for the 45\,m and
combined data, respectively.
For the remaining spectra, we failed to achieve a fit
because the weakest HF component of \tisolated\ 
whose intrinsic intensity is 0.111 times of that of the brightest \tbrightest\ 
did not have S/N-ratio higher than 5.\par

Figure \ref{fig:Nmaps} shows total column density maps of \NtwoH\ 
calculated from the 45\,m and combined data. 
To produce Figure \ref{fig:Nmaps}, we calculated \NNtwoH\ at
each pixel using Eq.(\ref{eq:NNtwoHfinal})
where we used \Tex, \taut, and intrinsic velocity width, \dVint,  
derived from the HFS analysis of the \NtwoH\ spectrum at a pixel.
The 45\,m \NNtwoH\ map shows that the core has the densest region at the center, 
and that the column density decreases outward.
The combined \NtwoH\ map shows that 
the extent of the densest central region is smaller than the 45\,m beam size.
The region has an approximate extent of 10\arcsec\
(corresponding to 0.01 pc) and \NNtwoH\ of $\gtrsim 10^{14}$ \cmq.
Notice that the 3\,mm source is shifted by 
6\arcsec\ to the east of the \NNtwoH\ peak.\par

To obtain fractional abundances of the probe molecules in $\S$\ref{ss:abundance},
we recalculate \NNtwoH\ with various beam sizes.
We smoothed the 45\,m data by 40\arcsec and 80\arcsec\ beams
for $r\leq 20\arcsec$ and 40\arcsec\ regions centered on the 3\,mm source.
Table \ref{tbl:Ncol} summarizes the calculated \NNtwoH\ 
and compares CCS and \amm\ column densities. 
We also reconstructed \NtwoH\ images from the combined visibilities
by giving the \sharc\ beam
as a synthesized beam size and calculated \NNtwoH\
inside the 3$\sigma$ level contour of the \um\ emission.\par

In addition, we analyzed the \NtwoH\ line data with the 45\,m telescope
to use the line as a temperature probe for the other molecules
because the \NtwoH\ HFS analysis gives rather accurate estimates
of \Tex\ and the spatial resolution for \NtwoH\ is the highest among our data.
We analyzed \NtwoH\ spectra inside the region enclosed by the half maximum, i.e., 
50\% level contour to the peak intensity and obtained mean \Tex\ of 
9.5$\pm$1.9 K (Table \ref{tbl:ResultsLine}).
We assume that the core has a temperature of 9.5 K and use
the temperature to calculate the column density of the other molecules.
Note that the adopted temperature is higher than the
range of 5 to 7 K for the ambient gas inferred from the \co\ (3--2) and (1--0) emission
($\S\ref{ss:res_outflows}$).

\subsubsection{\HtCOp\ (1--0)} 
\label{ss:col_h13co}

To estimate the column density of \HtCOp\  ($N_{\rm H^{13}CO^+}$),
we analyzed both the 45\,m telescope and combined data.
We firstly estimate optical depth (\tauhtcop) of the \HtCOp\ (1--0) 
line toward the core center from an antenna temperature ratio of
$T_{\rm A}^*$(H$^{13}$CO$^+$)/$T_{\rm A}^*$(HC$^{18}$O$^+$).
Here, we assume that the beam filling factors, \Tex, 
and the telescope beam efficiencies
are the same for the two isotope lines.
In addition, we assume that the solar abundance ratio 
$\frac{[\rm ^{13}C]}{[\rm ^{12}C]}
\frac{[\rm ^{16}O]}{[\rm ^{18}O]}$ of 5.5 
is applied to the $[\rm H^{13}CO^+]/[HC^{18}\rm O^+]$ ratio.
Using the peak \Tastar\ of 440 mK for the \HtCOp\ line in Table \ref{tbl:ResultsLine}
(\Tmb\ is converted into $T_{\rm A}^*$ by multiplying $\eta_{\rm mb}$ 
shown in Table \ref{tbl:sdobs}) 
and the 1.5$\sigma$ upper limit of 39 mK in \Tastar\ for the \HCeOp\ line 
(Table \ref{tbl:sdobs}),
we found that the \HtCOp\ line is optically thin 
($< 0.1$; Table \ref{tbl:ResultsLine}).
Given \tauhtcop\ of $< 0.1$ and \Tex\ of 9.5 K ($\S\ref{ss:col_n2h}$), we obtained
\NHtCOp\ from the 45\,m and combined data (Table \ref{tbl:Ncol}) 
by using Eqs.(\ref{eq:NHtCOpTmb}) and (\ref{eq:NHtCOpSnu}) 
(see Appendix \ref{ass:HtCOp}), 
where the \sharc\ beam was used as a synthesized beam to calculate \NHtCOp\ inside
the 3$\sigma$ contour of the \um\ emission.

\subsubsection{$C_3H_2$ $2_{12}-1_{01}$} 
\label{ss:col_CtHt}

For calculating \CtHt\ column density (\NCtHt) traced by the $2_{12}-1_{01}$ transition,
we followed the calculations by Benson, Caselli \& Myers (1998) (see their Appendix A2).
We observed \CtHt\ only toward the core center.
We estimated the optical depth using Eq.(17) by Benson et al. (1998) of 
$\tau_{\rm C_3H_2}=
\ln\left(\frac{T_{\rm ex}-T_{\rm mb}}{T_{\rm ex}-T_{\rm mb}-T_{\rm bg}}\right)$;~
\tauctht\ $ = 0.34$ for \Tmb\ of 1.1 K, 
assuming \Tex\ $ = 9.5$ K ($\S\ref{ss:col_n2h}$).
We derived \dVint\ (Table \ref{tbl:ResultsLine}) 
from the second order moment of the spectrum
because the line profile is not a single Gaussian (see Figure \ref{fig:spectra}).
Finally, a mean \CtHt\ column density is estimated to be 
$(2.4\pm 0.3)\times 10^{13}$ \cmq\ (Table \ref{tbl:Ncol}).\par

\subsubsection{CCS $4_3-3_2$} 
\label{ss:col_ccs}

In the similar manner as the \HtCOp\ analysis, 
we estimated the CCS column density (\Nccs) using Eq.(\ref{eq:Nccs}).
We obtain \tauccs\ of 0.94 (Table \ref{tbl:ResultsLine}) from 
a ratio of $T_{\rm A}^*$(CCS)/$T_{\rm A}^*$(CC$^{34}$S) $=$ 15.2 which is derived from 
the peak \Tastar\ of 560 mK for CCS (Table \ref{tbl:ResultsLine}) and 
a 1.5$\sigma$ upper limit of 37.5 mK for \ccsI\ (Table \ref{tbl:sdobs}).
Here we adopt the solar abundance ratio $[^{32}\rm S]/[^{34}\rm S]$ of 23.
Given \tauccs\ of 0.94 and \Tex\ of 9.5 K, 
we obtain \Nccs\ of $(1.5\pm 0.2)\times 10^{13}$ \cmq\ 
toward the core center with a $40\arcsec$ beam (Table \ref{tbl:Ncol}).\par

\subsubsection{\amm\ (1,1)} 
\label{ss:col_amm}

We present spectral line profiles of the \amm\ 
$(J, K)= (1,1)$ and (2,2) lines in Figure \ref{fig:ammspectra}.
The (1,1) spectrum consists of five groups of the lines.
The most intense emission in the (1,1) transition, 
the central group which contains eight of HF lines,  
has a peak \Tmb\ of 640 mK, whereas no (2,2) emission is seen 
above our 3$\sigma$ level detection threshold (\Tmb\ $=$ 210 mK).
Neither the (3,3) nor (4,4) emission was detected.\par

To estimate \amm\ column density ($N_{\rm NH_3}$), 
we followed the previous works, e.g., 
Winnewisser, Churchwell \& Walmsley (1979),
Ungerechts, Walmsley \& Winnewisser (1980),
Stutzki, Ungerechts \& Winnewisser (1982), and 
Ho \& Townes (1983).
We analyzed the spectra taken inside the 50\% level contour of the 
total intensity map (Figure \ref{fig:tiimaps}d).
In the similar manner as the \NtwoH\ HFS analysis, our ammonia HFS analysis gives 
\taut\ and \dVfwhm, and the LSR-velocity of the strongest HF component.
The brightest HF emission has optical depth of $\simeq 1.6$ toward the core center 
(Table \ref{tbl:ResultsLine}).
Here we assume that the beam filling factor is unity.
It is known that rotational temperature ($T_{\rm r,21}$) between the (1,1) and (2,2) 
transitions and the kinetic temperature ($T_{\rm k}$) agree well with each other
for high density ($\gtrsim 10^6$ \cmc) and cold ($\lesssim$ 10 K) gas 
(Tafalla et al. 2004).
We therefore assume $T_{\rm r,21}=T_{\rm k}$, and that 
\Tk\ is equal to \Tex(\NtwoH) of 9.5 K ($\S\ref{ss:col_n2h}$).
Now we calculate a mean column density of $(1.6\pm 0.6)\times 10^{14}$ \cmq\ 
for the \amm\ molecule in the (1,1) state inside the 50\% level contour 
of the total integrated intensity map by inverting Eq.(2) of Ungerechts et al. (1980)
We consequently obtained total \amm\ column density from the (1,1) column density
by Eq.(2) of Winnewisser et al. (1979) (Table \ref{tbl:Ncol}).\par

\subsection{Fractional Abundances of the Molecules in the Core}
\label{ss:abundance}

To calculate mass ($M_{\rm LTE}$) of the core from the column densities ($\S\ref{ss:rescolumn}$),
we need to find the fractional abundances (X$_{\rm m}$) of the probe molecules to H$_2$ molecules.
Such an abundance is known to vary under different physical and chemical conditions.
Bergin \& Langer (1997) theoretically predicted that all molecules are depleted
in some degree from gas phase for high densities of \nHmol\ $\gtrsim 10^6$ \cmc, 
regardless of dynamical evolutionary phases.
The depletion of sulfur-bearing molecules such as CS, SO, and CCS is 
particularly sensitive to the density 
increase because they tend to stick to both \wat\ and CO ice covered grains.
In contrast, nitrogen-bearing molecules of \amm\ and \NtwoH\ do not seem to be susceptible to 
freezing out on the grain surface (Bergin \& Langer 1997).
This prediction is confirmed by the recent observations toward starless cores, 
e.g., Tafalla et al. (2002) who showed that \NtwoH\ and \amm\ column densities 
correlate better with the dust column density than molecules such as CO and CS.
This characteristic, coupled with high angular resolution, makes 
\NtwoH\ ideal as a reference molecule.
In addition, we believe that the possible low-velocity compact outflow does not affect 
our discussion about molecular abundances, considering our findings from
radial density profile, position-velocity diagrams and velocity width maps of the core  
(described in $\S$\ref{ss:resdensity} and $\S$\ref{ss:resvelocity}).

\subsubsection{\NtwoH\ and \HtCOp}
\label{ss:Xnh}

The abundances of \NtwoH\ and \HtCOp\ are estimated through comparison
to the mean H$_2$ column density, $N$(H$_2$)$_{8\arcsec}$, derived from the \um\ continuum 
flux with the \sharc\ beam ($\S\ref{ss:rescont_dust}$).
Because the extent of the \um\ emission is approximately 4 times smaller than 
those of the \NtwoH\ and \HtCOp\ emitting regions,
such a comparison is possible only in 
the common innermost region of the core.
Comparing the $N$(H$_2$)$_{8\arcsec}$ with the \NtwoH\ and \HtCOp\ column densities with the
\sharc\ beam size (Table \ref{tbl:Ncol}), we obtained
X(\NtwoH)$_{8\arcsec}=$\XNtwoH\ and
X(\HtCOp)$_{8\arcsec}=$\XHtCOp\ (Table \ref{tbl:Xabs}).
Note that the errors in X$_{\rm m}$ are mostly due to the uncertainties in $\kappa_\nu$ 
($\S\ref{ss:rescont_dust}$).\par

We argue that \NtwoH\ in \gf\ is not depleted, whereas \HtCOp\ is depleted.
Our estimate of X(\NtwoH)$_{8\arcsec}$ is comparable to those in the following studies.
Womack et al. (1992) obtained a typical value of
X(\NtwoH )$=7\times 10^{-10}$ in the dark cloud cores TMC-1 and L134\,N 
(\thetahpbw\ $=70\arcsec$).
Benson et al. (1998) derived $(7\pm 4)\times 10^{-10}$ from a 
Haystack 37\,m ($22\arcsec$) survey towards 60 low-mass cores.
Caselli et al. (2002) reported $(3\pm 2)\times 10^{-10}$ 
from a FCRAO 13.6\,m ($54\arcsec$) mapping survey towards 34 dense cores.
Tafalla et al. (2002) reported the range of (0.75 -- 2.4)$\times 10^{-10}$ toward
5 starless cores using the FCRAO.
The consistency between our measurements with 8\arcsec\ beam and those with
22\arcsec, 54\arcsec, and 70\arcsec\ beams suggest that \NtwoH\ in \gf\ is not
depleted at the core center.
On the other hand, X(\HtCOp)$_{8\arcsec}$ in \gf\ is by
a factor of almost 6 smaller than the lowest abundances
in other similar cores measured with 67\arcsec\ beam
(Table 9 by Butner et al. 1995),
e.g., $5.6\times 10^{-11}$ for B\,335 (class 0) and 
$5.4\times 10^{-11}$ for L\,1495 (class I).
This fact suggests that \HtCOp\ in \gf\ is very likely to be depleted at the center, 
although we need to verify the radial variation of X(\HtCOp).

\subsubsection{\CtHt, CCS and \amm}
\label{ss:Xccsamm}

Using the X(\NtwoH)$_{8\arcsec}$ as a reference value, 
we estimated the abundances of \CtHt, CCS and \amm.
For this purpose, we assumed that \NtwoH\ in \gf\ does not have radial abundance variation.
Recall that Tafalla et al. (2002) reported that \NtwoH\ abundances are spatially constant 
in the 5 starless cores they surveyed.
Given the X(\NtwoH)$_{8\arcsec}$ and a column density ratio of N(\CtHt)/N(N$_2$H$^+$) 
with 20\arcsec\ beam (Table \ref{tbl:Ncol}),
we obtained X(\CtHt)$_{20\arcsec}$ of \XCtHt\ (Table \ref{tbl:Xabs}).
Similarly we obtained 
X(CCS)$_{40\arcsec}$ of \Xccs\ from an N(CCS)/N(\NtwoH) ratio for $r\leq 20\arcsec$, and
X(\amm)$_{80\arcsec}$ of \Xamm\ from an N(\amm)/N(\NtwoH) ratio for $r\leq 40\arcsec$.\par

The X(\CtHt)$_{20\arcsec}$ value in \gf\ would fall at the lower end of the
X(\CtHt) range, $(3.2 - 40)\times 10^{-10}$, obtained from the survey toward 
dense dark clouds with 54\arcsec\ beam (Cox, Walmsley \& G\"usten 1989),
although the beam sizes differ by a factor of $\sim$ 2.5.
The estimated X(CCS)$_{40\arcsec} =$ \Xccs\ is consistent with those of the starless cores 
measured with comparable beam sizes, i.e., 
$5.8\times 10^{-10}$ in TMC-1D and $4.9\times 10^{-10}$ in L\,1498 with
$45\arcsec - 53\arcsec$ beams (Wolkovitch et al. 1997), 
although smaller than $(8\pm 3)\times 10^{-9}$ with $\sim 52\arcsec$ beam 
in L\,1521\,F (Shinnaga et al. 2004).
On the other hand, the X(CCS)$_{40\arcsec}$ in \gf\ is found to be larger 
than those in the representative class 0 sources B\,1 and B\,335
($\simeq 9\times 10^{-11}$ and $\simeq 1.6\times 10^{-10}$, respectively)
reported in Lai \& Crutcher (2000) with $\simeq$ 30\arcsec\ BIMA beam.
Lastly, the X(\amm)$_{80\arcsec} =$ \Xamm\ in \gf\ is somewhat smaller than
the mean value derived from the large compilation of literature toward the cores 
in the Taurus and Ophiucus regions
($(4.0^{+2.3}_{-0.8})\times 10^{-8}$; Table B10 of Jijina et al. 1999), but is comparable
to the range of (0.4 -- 2.3)$\times 10^{-8}$ for the 5 starless cores surveyed by Tafalla et al. (2002).

Here we wish to discuss age of the \gf\ core in terms of chemical evolution.
Suzuki et al. (1992) analyzed their 45\,m telescope CCS survey data 
by developing a chemical evolutionary model, 
and concluded that CCS is the most abundant in the early evolutionary stage 
for a quiescent dark cloud core, 
whereas \amm\ is less abundant and becomes abundant in the later evolutionary stage.
Aikawa et al. (2001) advanced such a model for a dynamically collapsing core and
pointed out that \NtwoH\ shows similar chemical evolution as \amm. 
Comparing the X(\amm)$_{80\arcsec}$ and X(CCS)$_{40\arcsec}$ 
with Figure 16 in Suzuki et al. (1992) 
in conjunction with Figure 5 of Aikawa et al. (2001),
we argue that \gf\ core has experienced $10^5- 10^6$ years 
since the diffuse stage. Because of the large uncertainty in our abundance estimate,
we have limited our discussion to the 3 molecules whose abundance variation with time
are large enough to compare with our results.\par

\subsection{LTE Mass of the Core} 
\label{ss:Mlte}

Using the column densities of the \NtwoH, \HtCOp, CCS, and \amm\ lines ($\S\ref{ss:rescolumn}$)
and their fractional abundances ($\S\ref{ss:abundance}$), 
we now calculate the LTE mass (\MLTE) of the gas in the area ($\cal A$)
enclosed by the 50\% level contour of each total integrated intensity map 
of Figure \ref{fig:tiimaps} (see Table \ref{tbl:Mcore}).
The mass is given by
$M_{\rm LTE} = \mu m_{\rm H}\cdot \frac{N_{\rm m}}{X_{\rm m}}\cdot {\cal A}$
where $\mu$ is the mean molecular weight 
(2.33 for [He]$=0.1$[H]) and $m_{\rm H}$ is the atomic mass of hydrogen.
The obtained LTE masses range from 1.2 to 4.5 \Msun.

\subsection{Radial Column Density Profile of the Core}
\label{ss:resdensity}

We derive radial column density profile of \gf\ core
using the \NtwoH\ column density maps.
Recall that the \NtwoH\ in \gf\ is unlikely to be depleted over the full extent of the core, 
whereas the \HtCOp\ may be depleted ($\S\ref{ss:Xnh}$).
We assume that the core is spherical.
To azimuthally average the column density maps, 
we take the \NtwoH\ emission peak as the center.
Figure \ref{fig:radprofile} represents the resultant column density profiles.
The profiles can be well fitted by a power-law function of 
$N(r)=N_0(\frac{r}{r_0})^{-p}$
where $p$ is the power-law index, and $N_0$ is 
the column density at $r = r_0 = 1\arcsec$.
By minimizing the $\chi^2$ values, 
we obtained the best-fit parameters of 
$p=1.0\pm 0.1$ and $N_0=\left(7.3\pm 1.1\right)\times 10^{14}$ \cmq\ for the combined data, and
$p=1.1\pm 0.2$ and $N_0=\left(8.5\pm 2.5\right)\times 10^{14}$ \cmq\ for the 45\,m telescope data.
The two fitting results show good consistency within uncertainties, 
indicating that a negligible amount of the dense gas has been evacuated
by the possible compact outflow in the innermost region of $r\lesssim$ 3\arcsec.\par

\subsection{Velocity Structure of the Core} 
\label{ss:resvelocity}

\subsubsection{Overall Velocity Structure} 
\label{ss:pv}

To examine the velocity structure of \gf\ core,
we made position-velocity (PV) diagrams using the 45\,m 
\isolated\ and \HtCOp\ (1--0) line data (Figure \ref{fig:pv}).
We also overlaid PV diagrams made from the combined \HtCOp\ data set.
Since the core is elongated 
along the NE-SW (Figures \ref{fig:tiimaps}, \ref{fig:NROchmaps} and \ref{fig:Nmaps}),
we select the two cutting axes at P.A. of 45\degr and 135\degr,
passing through the position of the 3\,mm source
as the major and minor axes, respectively. 
Along the major axis, the 45\,m data show
a weak velocity gradient from NE toward SW
in both the molecular lines.
The gradient along the major axis is an order of 1 \kms\ pc$^{-1}$,
comparable to the typical value measured in the Taurus dark cloud cores (Goodman et al. 1993).
If the gradient represents the rigid-body-like rotation of the core, 
it is converted to an angular velocity of 
$\Lambda \simeq 3\times 10^{-14}$ s$^{-1}$.
On the other hand, the minor axis plots do not show 
a well-defined velocity gradient commonly seen in the two lines,
although the \HtCOp\ diagram appears to show an SE-NW gradient
if we concentrate on the lowest two contours.\par

Another important finding from Figure \ref{fig:pv} is the presence of 
broader line width region at the core center, prominently seen for the \HtCOp.
The blueshifted gas is seen up to $\simeq$ 0.6 \kms\ from the \Vsys\ at the central 
region of $r\lesssim 5\arcsec$ (corresponding to 1000 AU), 
whereas the redshifted counterpart about $\simeq$ 0.3 \kms.
The presence of such blueshifted emission and the weak 
redshifted counterpart at the core center (see also Figure \ref{fig:spectra})
suggests that gas is infalling where redshifted gas
is obscured by the foreground cold gas along the line-of-sight
(e.g., Walker et al. 1986). Notice that the line broadening at the core center is unlikely 
to be caused by the possible compact outflow because the red lobe is more intense than 
the blue one (see Figure \ref{fig:COBluRed}), inconsistent with the above results.\par

\subsubsection{Velocity Width}
\label{ss:widthmap}

We firstly estimate intrinsic velocity width (\dVint) toward the core center
by analyzing the \NtwoH, \HtCOp, \CtHt, CCS, and \amm\ lines (Table \ref{tbl:ResultsLine}).
Note that \dVint\ is the velocity width corrected for line broadening 
due to instrumental velocity resolution and optical depth of the line (Phillips et al. 1979).
Both corrections are made for the \CtHt\ and CCS lines. 
The latter correction is not made for \NtwoH\ and \amm\ data
because HFS analysis obtains the optical-depth-free \dVfwhm\ (see Appendix B), 
and not for the \HtCOp\ data because it is optically thin.
The obtained \dVint\ toward the center is between 0.35 and 0.61 \kms.
This range is 2$\sim$3.5 times larger than the isothermal sound speed 
(\Cs) of the ambient gas of 0.18 \kms\ at \Tk\ $=$ 9.5 K ($\S\ref{ss:col_n2h}$).\par

Subsequently, we analyzed velocity width structure of the core.
Figure \ref{fig:dVint} shows the \dVint\ maps made by spectral moment analysis 
of the 45\,m telescope \isolated\ and \HtCOp\ (1--0) line data.
We adopt the moment method to keep a consistency in the analysis between the two lines.
To make the maps, we made spectra from the cube data at each point, 
then searched for a peak LSR-velocity in each spectrum to determine 
two velocity ranges from the peak velocity toward both the blue- and redshifted sides 
where the intensities are above the 1.5$\sigma$ level. 
We calculated the second-order spectral moments
over the velocity ranges defined for the individual spectra.
For the \NtwoH\ map (Figure \ref{fig:dVint}b), 
we corrected the line broadening effect due to optical depth.
We used  Eq.(3) in Phillips et al. (1979) by substituting
$\tau_{\rm p}$ of Eq.(3) with $\tau_{\rm tot}\times s$ 
where $s$ is the intrinsic relative intensity of 
the \tisolated\ transition (0.111; Cazzoli et al. 1986).
The maximum line broadening at the most opaque ($\tau_{\rm tot}\times 0.111=$ 1.5) 
position is 1.3 times the width for $\tau_{\rm p}\rightarrow 0$.
Considering the thermal noise and the velocity resolutions, 
we estimate uncertainties of 0.11 and 0.12 \kms\ in \dVint\
for \NtwoH\ and \HtCOp, respectively.
As seen in PV diagrams ($\S\ref{ss:pv}$), 
both the \NtwoH\ and \HtCOp\ maps show a clear increase of \dVint\ toward the core center.
The peak of the broadening does not fall onto the core center.
It also should be noted that 
\dVint\ of the core gas exceeds \Cs\ of 0.18 \kms\ 
over the core (see also Table \ref{tbl:ResultsLine}).
Furthermore, comparing Figures \ref{fig:COBluRed} and \ref{fig:dVint}, 
one cannot discern any clear effects such as an evacuated shell by 
the possible compact outflow in Figure \ref{fig:COBluRed};
the \co\ (3--2) condensations are only observed toward 
the 3\,mm continuum source.\par

The above analysis clearly shows that 
there exist non-thermal motions in the core.
We estimate non-thermal velocity width (\dVnth) by
subtracting the thermal velocity width (\dVthm) from \dVint:~
$(\Delta v_{\rm int})^2=
(\Delta v_{\rm thm})^2+(\Delta v_{\rm nth})^2$.
Here the thermal contribution is given by 
\dVthm\ $=\sqrt{\frac{8(\ln 2) kT_{\rm k}}{m_{\rm mol}}}$
where $k$ is the Boltzmann constant, and $m_{\rm mol}$ is the
molecular weight:~
29 amu for \NtwoH, 
30 amu for \HtCOp,
56 amu for CCS, and 17 amu for \amm .
Assuming that \Tk\ is equal to \Tex\ of 9.5 K ($\S\ref{ss:col_n2h}$) and 
that \Tk\ is constant over the entire core, 
\dVthm\ is calculated to be
0.12 \kms\ for \NtwoH,
0.12 \kms\ for \HtCOp,
0.09 \kms\ for CCS, and
0.16 \kms\ for \amm\ lines.
This leads to mean non-thermal velocity widths over the core of
\meandVnth\ $=$ 
0.21, 
0.58,
0.40, and 
0.46 \kms\ for the 
\NtwoH, 
\HtCOp, 
CCS, and 
\amm\ lines, respectively, 
suggesting the presence of supersonic gas motions. In obtaining \meandVnth, 
we used the \meandVint\ values in Table \ref{tbl:ResultsLine}.\par

\section{Discussion}
\label{s:Discussion}
\subsection{Evolutionary Stage of the GF\,9-2 Core: 
A Very Young Protostar Before Formation of a Large-Scale Molecular Outflow}
\label{ss:evl_flow}

With its low luminosity ($\simeq 0.3$ \Lsun) and cold SED (\Tbol\ $\leq 20$ K; 
Wiesemeyer et al. 1997),
we believe that the protostar embedded in \gf\ core is at an extremely
early phase of low-mass star formation, 
evidenced by the absence of a large-scale high-velocity molecular outflow,
the possible detection of a compact low-velocity outflow, and
the detection of \wat\ masers. 
We estimate the protostar's age from dynamical age (\tdyn) of the possible
compact low-velocity outflow ($\S\ref{ss:res_outflows}$).
The terminal velocity ($V_{\rm t}$) of the spectra would give 
the line-of-sight component of the outflow velocity of
$|V_{\rm t}-V_{\rm sys}|=V_{\rm flow}\cos i$ where $i$ is the inclination angle of the 
outflow axis to the line-of-sight.
Given the projected lobe length of $l_{\rm lobe}\sin i = 1,1000$ and 6600 AU 
and $V_{\rm flow}\cos i \simeq$ 3.3 and 3.1 \kms\ ($\S$\ref{ss:res_outflows}), 
we have $t_{\rm dyn}=l_{\rm lobe}/V_{\rm flow}\simeq 1.6\times 10^4$ and $1.0\times 10^4$ years 
with $i=$ 45\degr\ for the blue- and redshifted \co\ (3--2) lobes, respectively.
These dynamical ages are comparable to those 
for the extremely young class 0 sources,
e.g., (0.6--3)$\times 10^4$ years for VLA\,1623 and (0.8--2)$\times 10^4$ years for IRAM\,04191
(Table 1 in Andr\'e, Motte \& Backman 1999, and references therein).\par

The youth of \gf\ is demonstrated in Figure \ref{fig:gfevl} where
we present a plot of CO outflow momentum rate (\Fco) normalized by the bolometric luminosity
(\FcocLbol) vs. \Tbol\ for nearby ($d\leq 350$ pc) low-mass (\Lbol\ $\leq 50$ \Lsun) YSOs.
We employ \FcocLbol\
because molecular outflows in low-mass protostars are likely to be momentum-driven
(e.g., Cabrit \& Bertout 1992; Richer et al. 2000, and references therein).
The temperature \Tbol\ is believed to trace the evolutionary stage of YSOs (Myers \& Ladd 1993).
We calculate 3$\sigma$ upper limit of \FcocLbol\ 
$= \frac{M_{\rm lobe}V_{\rm flow}^2}{l_{\rm lobe}}\cdot \frac{c}{L_{\rm bol}}$
$\simeq$ 9 and 20 for the \co\ (3--2) blue- and red lobes, respectively, 
with $i=$ 45\degr.
Figure \ref{fig:gfevl} illustrates that ``class 0'' is the most powerful outflow phase 
and ``class I'' is the decline phase (e.g., Moriarty-Schieven et al. 1994).
The most remarkable result is that \gf\ shows the lowest outflow activity, 
among the coldest low-mass class 0 sources, 
although there exists enough circumstellar material ($\S$\ref{ss:Mlte}) 
to be dragged by a possible protostellar jet to develop a large-scale molecular outflow.
Note that the prototypical class 0 sources of VLA\,1623 (AWB93)
and IRAM\,04191 (Andr\'e et al. 1999) have powerful
CO outflows, despite that their
bolometric luminosities are comparable to that of \gf:
1 \Lsun\ for VLA\,1623 and 0.15 \Lsun\ for IRAM\,04191.
We thus believe that the low luminosity of \gf\ is not responsible for
the lowest outflow activity.  Consequently, it is very likely that 
the protostar of our interest is the youngest class 0 protostar ever identified
and is just at the beginning of the outflow launch.\par

\subsection{The Initial Conditions for Gravitational Collapse of a Low-Mass Star Forming Core}
\label{ss:ICofSF}

\subsubsection{Stability of the \gf\ Core}
\label{ss:stab}

If we interpret that the non-thermal velocity width obtained in \gf\ core 
($\S$\ref{ss:widthmap}) is attributed to turbulent motions, 
we expect that the turbulence can help support the core against its self-gravity.  
To test this hypothesis, we compare the LTE mass of the core with virial mass 
\Mvir\ including the turbulence.  
Assuming a spherical system, the virial mass is given by 
$M_{\rm vir}\simeq R_{\rm eff}\langle\Delta v_{\rm int}\rangle^2/G$
where \meandVint\ includes the non-thermal width, i.e., the turbulence.
Given \meandVint\ and \Reff\ values (Tables \ref{tbl:ResultsLine} and \ref{tbl:Mcore}), 
we found that \Mvir\ ranges from 0.39 to 2.9 \Msun (case I in Table \ref{tbl:Mcore}).
Since \Mvir\ $\simeq$ \MLTE\ within the uncertainties,
the core is most likely to be in virial equilibrium.
However, it should be noted that this hypothesis does not reconcile with 
the line broadening toward the core center (Figure \ref{fig:dVint}).
Moreover, as far as we know, no cores which are supported by turbulence exhibit 
a \RadpSimSoltn\ density profiles (see Figure \ref{fig:radprofile}).
Therefore, we conclude that the non-thermal line width 
is not caused by turbulence.\par

Alternatively, if we consider that infall causes the non-thermal line width, 
that is, if \gf\ core is gravitationally unstable, the drawbacks in the previous 
hypothesis would be solved. The presence of infalling motion
is supported by the following facts:~ 
(1) the presence of the blueshifted gas and the less prominent redshifted counterpart
in the central $r\lesssim$ 1000 AU ($\S$\ref{ss:pv}),
(2) the spectroscopic signature of infall, i.e., 
the ``blue-skewed'' profile observed in the optically thick \isolated\ line (Figure \ref{fig:spectra}), and
(3) the detection of the \wat\ masers, a clear signpost of star formation, 
naturally leading to the presence of accretion flow onto the star-disk system.
We, however, point out a drawback in the above (2); the optically thin
\HtCOp\ emission does not show a single Gaussian profile expected from 
the infall interpretation, while the CCS does.
In this interpretation, \Mvir\ should be estimated only from the thermal motion as 
\Mvir\ $=\frac{R_{\rm eff}}{G}\frac{kT_{\rm k}}{\mu m_{\rm H}}$.
The estimated \Mvir\ from the 4 lines takes a range
between 0.078 and 0.25 \Msun\ (case II in Table \ref{tbl:Mcore})
which is one order of magnitude smaller than \MLTE.
The small \Mvir\ is consistent with our interpretation that 
the core is under gravitational collapse.
As discussed in $\S\ref{ss:widthmap}$, 
the core would be rotating with $\Lambda\simeq 10^{-14}$ s$^{-1}$.
The collapse cannot be supposed by the centrifugal force due to the 
rotation because the rotational energy is less than a few percent of total 
energy required to support the core against its self-gravity.
For such an unstable core, 
the gas is expected to be more perturbed toward the center, 
and the velocity width should increase inward as a result.
In fact, Figure \ref{fig:dVint} clearly indicates that \dVint\ in \gf\
increases up to $\simeq 0.6$ \kms\ toward the center.\par

\subsubsection{Initial Conditions for the Gravitational Collapse of the Core}
\label{ss:IC}

To bring out the initial conditions for the gravitational collapse in \gf\ core, 
we examine the density and velocity structures by comparing with 
some key characteristics of the Shu (1977) and Larson-Penston (Larson 1969; Penston 1969) solutions.
The radial H$_2$ column density profile of the core is given by 
$(7.7\pm 4.1)\times 10^{23}\left(\frac{{\rm X(N_2H^+)}}{9.5\times 10^{-10}}\right )^{-1}
\left(\frac{r}{200 \rm AU}\right )^{-(1.0\pm 0.1)}$ \cmq\ for 0.001 $< \frac{r}{\rm pc} <$ 0.05 
from the combined data and 
$(8.9\pm 4.5)\times 10^{23}\left(\frac{{\rm X(N_2H^+)}}{9.5\times 10^{-10}}\right )^{-1}
\left(\frac{r}{200 \rm AU}\right )^{-(1.1\pm 0.2)}$ \cmq\ for 0.005 $< \frac{r}{\rm pc} <$ 0.08
from the 45\,m data (Figure \ref{fig:radprofile}).
It should be noted that the dense \NtwoH\ gas seem to obey the power-law 
profile with the index of $-1$ even at the innermost region of $r\lesssim$ 1000 AU
where a possible compact outflow might begin to destroy the core.
The power-law indices of $-1.0$ and $-1.1$, 
converted into $-2.0$ and $-2.1$ for the volume density, 
well agree with those in the two solutions after the formation of the protostar 
(Shu 1977; Whitworth \& Summers 1985).
The best-fit H$_2$ column densities at $r =$ 200 AU fall 
between the two extreme predictions at $T =$ 9.5 K:
$(2.8\pm 1.5)\times 10^{23}$ \cmq\ for the Shu and 
$(13\pm 7)\times 10^{23}$ \cmq\ for the Larson-Penston solutions.
Here we used a relation of $N_0({\rm H_2}) =$ 1.0 and 
$4.4\times\frac{c_{\rm s}^2}{2Gr_0}\cdot\frac{{\rm X(N_2H^+)}}{\mu m_{\rm H}}$ 
for the Shu's and Larson-Penston's cases, respectively 
(Hunter 1977; Shu 1977; Foster \& Chevalier 1993).
Because of the abundance uncertainty, however, 
we cannot conclude which solution describes 
well the collapse in \gf\ core.

Contrary to the above discussion based on the density profile, 
the velocity width maps (Figure \ref{fig:dVint}) 
help to discriminate the two theoretical scenarios for the collapse of \gf\ core.
Notice that the influence of the compact outflow on the velocity
width cannot be discerned toward the 3\,mm source.
Figure \ref{fig:dVint} shows that the velocity width decreases outward, 
but remains $\simeq$ 0.3 \kms, which is twice the ambient \Cs, 
even in the outermost region of $r\simeq 50\arcsec$ (corresponding to 0.05 pc) where
the density profile shows \RadpSimSoltn\ (Figure \ref{fig:radprofile}).  
Recall that \RadpSimSoltn\ profile characterizes outer part of a collapsing core.
These facts clearly indicate that the extended Larson-Penston solution for $t>$ 0 is 
more likely to be the case because it requires the infalling state 
when a protostar is about to be born
($v(r) \sim -3.28 c_{\rm s}$ at $t \rightarrow -0$; Hunter 1977),
whereas the Shu solution requires a static state ($v(r) =$ 0 at $t =$ 0).
It is, therefore, highly probable that the gravitational collapse of \gf\ core is well 
described by the runaway collapse scenario.\par

Subsequently, we estimate the age of the \gf\ protostar (\tpstar) 
based on the lack of the density profile of \RadpInf\ down to the innermost region of 
$r\simeq 3\arcsec$ (see Figure \ref{fig:radprofile}).
Recall that \RadpInf\ profile is expected for a central free-fall region
which appears in both of the solutions at $t >$ 0. 
The non-detection of a free-fall density profile indicates 
that the central free-fall region, propagating outward with supersonic velocity, 
has not yet developed up to a radius ($r_{\rm ff}$) of $\simeq$ 600 AU,
suggestive of the youth of the \gf\ protostar.
However, we do not exclude the possibility that 
the absence of a detection of the \RadpInf\ profile at the center 
could be caused by the formation of binary (described in $\S\ref{ss:pbinary}$)
whose projected separation is $\simeq$ 1200 AU.
Using Eqs. (9) and (10) in Ogino, Tomisaka \& Nakamura (1999), 
we estimate \tpstar\ to be less than
\tpstar\ $\simeq 5000\left(\frac{T}{\rm 9.5~K}\right)^{-\frac{1}{2}}$ years 
since the formation of the central protostar for the Larson-Penston solution.
Although the age estimate is a factor of 2--3 smaller than the outflow dynamical age of 
\tdyn\ $\simeq (1.0-1.6)\times 10^4$ years ($\S$\ref{ss:evl_flow}), we argue
that the central protostar has just begun to exhibit its jet activity
traced by the \wat\ masers ($\S$\ref{ss:evl_flow}).\par

Consequently, we propose that \gf\ core has started its gravitational collapse with 
initially unstable state.
Here, the term {\it initial} means $t= -t_{\rm ff}$
where \tff\ is free-fall time of the core (Larson 1969; Penston 1969). 
For \gf, we calculate \tff\ $\simeq 2\times 10^{5}$ years. 
We therefore believe that the age of the core, 
\tcore\ $=$ \tff\ $+$ \tpstar, would be $\simeq 2\times 10^{5}$ years, 
which has a consistency with the time scale of chemical evolution 
($\S$\ref{ss:Xccsamm}).\par

\subsubsection{Infall Velocity and Mass Accretion Rate in the Core}
\label{ss:Mdot}

Assuming that \dVint\ represents a velocity difference of blue- and redshifted infalling gas
along the line-of-sight, let us discuss infall velocity (\Vinf) of the core gas.
We found that \Vinf\ varies from 0.2 to 0.4 \kms\ using a relation of
\Vinf\ $\simeq \frac{1}{2}|v_{\rm inf, red}-v_{\rm inf, blue}|/\cos i\simeq$
$\frac{\dVint}{2\cos i}$ for the \dVint\ range of (0.28 -- 0.63) \kms\ over the core
(see Figure \ref{fig:dVcenter}) and $i=$ 45\degr.
We now calculate the mass infall rate from
$\frac{dM}{dt} = 4\pi R_{\rm eff}^2 \mu m_{\rm H}\cdot n_{\rm H_2}(R_{\rm eff})\cdot v_{\rm inf}$ 
$\simeq 2\pi R_{\rm eff}\cdot \frac{\mu m_{\rm H}\cdot N_{\rm N_2H^+}(R_{\rm eff})}{{\rm X(N_2H^+})}\cdot v_{\rm inf}$
for a sphere with the radius of \Reff.
Here $n_{\rm H_2}(R_{\rm eff})$ is a volume density of a thin shell whose radius is \Reff\
and is approximated as $\frac{N_{\rm N_2H^+}(R_{\rm eff})}{2R_{\rm eff}\cdot{\rm X(N_2H^+)}}$.
Given \Vinf\ $=$ 0.29 \kms\ at $r \simeq$ \Reff\ 
from the \NtwoH\ \dVint\ map (Figure \ref{fig:dVint}), 
we have $\frac{dM}{dt}\simeq 3\times 10^{-5}$ \Msun\ year$^{-1}$ 
with uncertainty of a factor 1.6 for 
\Reff\ in Table \ref{tbl:Mcore}, 
the column density profile ($\S\ref{ss:resdensity}$), 
and X$_{\rm m}$ in Table \ref{tbl:Xabs}. 
The inferred mass accretion rate falls within the range between the predicted values of 
$1.5\times 10^{-6}$ and $ 7.4\times 10^{-5}$ \Msun\ year$^{-1}$ for the Shu and Larson-Penston
solutions, respectively, at 10 K.
These accretion rates are calculated by $m_0 c_{\rm iso}^3/G$ where $m_0$ is
the non-dimensional factor of 0.975 for the Shu and 
46.92 for the Larson-Penston solutions (Hunter 1977).
Our small mass accretion rate, compared to that of the Larson-Penston's solution,
is attributed to both the smaller column density and infall velocity.

\subsection{Is a Protobinary System Forming ?}
\label{ss:pbinary}

We found that the position of 3\,mm continuum source is displaced by 
$ \simeq 6\arcsec$ to the east of the \NtwoH\ column density peak, 
which would be explained by the presence of a protobinary system.
In this interpretation, the 3\,mm source represents a protostar and 
the \NtwoH\ peak a preprotostellar condensation.
This is because the \NtwoH\ condensation is deeply embedded in the extended \um\ emission
(Figure \ref{fig:Nmaps}b), but is not accompanied with a compact 3\,mm source.
To investigate the nature of the binary members,
we made a \dVint\ map from the combined \HtCOp\ data shown in Figure \ref{fig:dVcenter}
as the line is expected to be optically thin.
In $\S\ref{ss:rescont_dust}$, 
we asserted that the 3\,mm continuum emission would represent circumstellar 
material around a protostar.
The line broadening toward the 3\,mm source
seen in Figures \ref{fig:pv} and \ref{fig:dVcenter}
suggests the local enhancement of mass accretion, 
leading to the existence of a protostar.
On the other hand, the \NtwoH\ peak (white contour) seems to be a gas condensation 
or a preprotostellar core rather than an apparent peak caused by local abundance enhancement
because X(\NtwoH)$_{8\arcsec}$ inferred over the 3$\sigma$ level contour of \um\ 
emission is comparable 
to those obtained with low spatial-resolution studies ($\S\ref{ss:Xnh}$).
It should be noted that the \HtCOp\ emission has a secondary peak 
near the \NtwoH\ condensation as seen in the OVRO maps (Figure \ref{fig:hikaku}).
Furthermore, the combined \NtwoH\ map (Figure \ref{fig:hikaku})
does not show a shell-like structure which is expected to be seen if \NtwoH\ disappears
from the gas phase at the core center
(e.g., Kuiper, Velusamy \& Langer 1996; Belloche \& Andr\'e 2004).
We therefore conclude that the positional displacement of \NtwoH\ peak 
is not caused by chemical evolution, 
but the \NtwoH\ peak represents a preprotostellar condensation, 
making it younger than the eastern object.
The detection of the two objects reinforces the assertion that
we witness a forming binary system with a separation of $\simeq$ 1200 AU.

The masses of the circumstellar material ($M_{\rm cm}$) 
for each binary member would be an order of 0.1 \Msun. 
For the eastern object, the 3\,mm continuum flux density ($\S$\ref{ss:rescont_dust})
integrated inside its 3$\sigma$ level contour (see Figure \ref{fig:dVcenter})
implies that the gas plus dust mass ranges from
$\simeq$ 0.1 (case I in $\S\ref{ss:rescont_dust}$) to 0.2 \Msun\ (case II). 
For the western object, we estimate (0.1 -- 0.3) \Msun\ from the \NtwoH\ intensity 
integrated inside the white contour, 
whose extent is similar to that of the 3\,mm source (Figure \ref{fig:dVcenter}).
The inferred $M_{\rm cm}$ seem reasonable for an amount of circumstellar material
with 100 AU scale around a newly formed protostar
at the very early accretion phase because of 
$M_{\rm cm} \simeq M_{\rm protostar} = \frac{dM}{dt}\times t_{\rm protostar} \simeq 0.15$ \Msun\ 
($\S$\ref{ss:IC} and \ref{ss:Mdot}).
In this context, the EW elongated structure at the core center seen in the 
45\,m maps (Figure \ref{fig:hikaku}) may be interpreted as
a 1000 AU scale circumbinary disk-like structure 
because the size would be smaller than the extent of the
protostellar envelope traced by the \um\ continuum emission 
(6600$\times$4500 AU; $\S$\ref{ss:rescont_dust}).
It is impossible to uniquely define spatial extent of the disk-like structure.
However, if we arbitrarily define the disk-like structure by a region enclosed 
by the 18 or $21\sigma$ level contours of the combined \HtCOp\ map 
(Figure \ref{fig:dVcenter}),
mass of the circumbinary disk-like structure (\Mcbd) 
calculated from the \NtwoH\ data would range
(0.2 -- 0.5) \Msun\ with an approximate effective radius of $\simeq$ 2000 AU.
Considering the uncertainties in spatial extents and
in $\kappa_0$, $\beta$ ($\S$\ref{ss:rescont_dust}) and X(\NtwoH) ($\S$\ref{ss:Xnh}), 
the mass of disk-like structure is likely to be plausible
to be hosted in the (0.4 -- 0.8) \Msun\ envelope ($\S$\ref{ss:rescont_dust}) 
and to host the two binary members with 
the 0.1 \Msun\ order circumstellar material.
The mass distribution of 
$M_{\rm core} > M_{\rm env} \gtrsim M_{\rm cbd} \gtrsim M_{\rm cm}$
does not contradict with our hypothesis that the central protostar(s) is
at extremely early evolutionary phase.\par

If a circumbinary disk exists, the radius of disk-like structure would be physically 
related to the rotation of the natal core, described in $\S\ref{ss:pv}$.  
Although the axis for the large-scale rotation of the core (P.A.$\simeq$ 135\degr) 
is not perpendicular to the major axis of the 
disk-like structure (P.A.$\simeq$ 90\degr), 
a rough estimate of radius of the disk-like structure would be given by the centrifugal radius, \Rcent,
calculated from the specific angular moment of the core gas.  
With a relation of $R_{\rm cent}=(4R_{\rm eff}^4\Lambda^2)/(GM_{\rm cbd})$,
we obtained \Rcent\ $\simeq 900$ AU for 
$\Lambda$ of $3\times 10^{-14}$ s$^{-1}$ ($\S\ref{ss:pv}$) and
\Reff\ in Table \ref{tbl:Mcore}.
The centrifugal radius is roughly twice the radius of disk-like structure, 
probably supporting our interpretation.  
Lastly, let us estimate the Toomre's $Q$ parameter to examine whether or not 
gravitational fragmentation of the disk-like structure led to 
the formation of the binary.
The $Q$ parameter is given by 
$\frac{\Lambda c_{\rm s}}{\pi G \Sigma}$ 
for Keplerian rotation.
Here $\Sigma$ is the surface density of the disk-like structure 
given by $M_{\rm cbd}/{\cal A}_{\rm cbd}$;
an area of ${\cal A}_{\rm cbd}$ $\simeq 4\times 10^{33}$ cm$^2$ is
calculated for the region described above.
We obtained $Q\simeq 0.01
\left (\frac{\Lambda}{3\times 10^{-14} \rm ~s^{-1}}\right )
\left (\frac{T}{\rm 10~ K}\right )^{\frac{1}{2}}
\left (\frac{M_{\rm cbd}}{0.4 M_{\odot}}\right )^{-1}
\left (\frac{{\cal A}_{\rm cbd}}{4\times 10^{33}~ \rm cm^2}\right ) $.
Considering the uncertainties in the input parameters, 
it is certain that the $Q$ value is less than unity.  
We, therefore, suggest that the binary system is likely to be forming 
in the core center 
due to the gravitational fragmentation of the circumbinary disk-like structure, 
whose radius appears to be determined by the rotation of the core.

\section{Summary and Future Works}

The main findings of this work and future considerations are summarized as follows.

\begin{enumerate}

\item Our high density gas observations identified a molecular cloud core with 
a size of $\simeq$ 0.08 pc toward \gf.  
In the core center, a protostellar envelope with a size of $\simeq$ 4500 AU
is identified through \um\ continuum emission.
The mass of the core and the envelope are $\simeq$ 3 and $\simeq$ 0.6 \Msun, respectively.  
Obtaining fractional abundances of the probe molecules, we
propose that \gf\ core should be at an early phase of chemical evolution, 
although the radial variations of the abundances
should be revealed to accurately assess the chemical evolutionary stage.

\item Our deep search for CO wing emission revealed that 
the protostar in \gf\ has not developed a large-scale high-velocity 
``jet-like'' outflow.
Instead, a compact low-velocity outflow may be formed at the core center.
Thus, \gf\ is very likely to be at an extremely early stage in 
low-mass star formation before arriving its most active outflow phase.
To test the conclusion, we need to conduct high angular resolution 
and deeper search for compact outflow(s) in higher-$J$ transitions of CO.

\item The radial column density profile of the core can be well fitted by 
a power-law form of \RadpSimSoltn\ for 0.003 $< r/{\rm pc} <$ 0.08 region.
The power-law index of $-2$ agrees with the expectation for an outer part of 
gravitationally collapsing core.
On the other hand, based on the absence of a \RadpInf\ profile, 
we found that a central free-fall region has not developed in a region larger
than $r\simeq 600$ AU.

\item The PV analysis identified a large-scale velocity gradient with an
order of 1 \kms\ pc$^{-1}$ along the major axis of the core.
The gradient is interpreted as a rigid-body-like rotation 
with an angular velocity of $\simeq 3\times 10^{-14}$ s$^{-1}$,
comparable to those measured in other similar cores.  
We found that the gas is supersonic all over the core
and that the velocity width is enhanced up to $\simeq$ 0.6 \kms\ in the central 
1000 AU toward the 3\,mm source.
The supersonic velocity width and the \RadpSimSoltn\ profile 
are explained by infall in the extension of the runaway collapse scenario 
proposed by Larson and Penston for $t>0$.
The current mass accretion rate of $\simeq 3\times 10^{-5}$ \Msun\ year$^{-1}$ 
is estimated from \Vinf\ $\simeq$ 0.3 \kms\ at $r \simeq$ 7000 AU. 
We discuss the likelihood that \gf\ core has been undergoing gravitational 
collapse $\lesssim$ 5000 years, starting from initially unstable conditions.

\item We discovered a potential protobinary system with a projected separation 
of $\simeq$ 1200 AU embedded in a circumbinary disk-like structure with 
$\simeq$ 2000 AU radius at the core center. 
The binary consists of a very young protostar and a preprotostellar condensation.
The circumbinary material mass is on the order of 0.1 \Msun.
Taking account of the centrifugal radius 
calculated from the core rotation and the Toomre's $Q$ value for the disk, 
we suggest that the binary formation may have been initiated by gravitational 
fragmentation of the circumbinary disk-like structure.
Further studies are needed using high-angular-resolution deep imaging 
in submm and FIR continuum emission.

\end{enumerate}


\acknowledgments

The authors sincerely acknowledge an anonymous referee whose comments significantly
improved the quality of the paper.
R.S.F. acknowledges A. I. Sargent and J. M. Carpenter for critical readings
of the manuscript and continuous encouragement.
The authors thank H. Yoshida for providing the telescope time at CSO to obtain the 
\co\ (3--2) data, S. Takakuwa, J. Lamb, D. Woody, K. Sunada, and S. Okumura 
for fruitful discussion about data combining of single-dish telescope and interferometer,
H. Wiesemeyer for information on \gf\ including the distance and the luminosity, and
L. Testi for his early contribution to this study.
R. S. F. also appreciates discussion with 
A. Wootten, R. Cesaroni, E. van Dishoeck, H. Butner, G. Blake, 
T. Hanawa, T. Matsumoto, Y. Aikawa, K. Tachihara,
T. Velusamy, and D. Ward-Thompson.
R. S. F. and H. S. thank M. Yang, A. Kov\'acas, C. Borys and A. C. A. Boogert
for their support at the \sharc\ observations and data reduction.
The authors gratefully acknowledge all the staff at 
OVRO, NRO, CSO, VLA, and the GILDAS software group at IRAM.
Research at the Owens Valley Radio Observatory
is supported by the National Science Foundation through NSF grant AST 02-28955.

\appendix

\section{Combining the OVRO mm-array and Nobeyama 45\,m Telescope Data}
\label{as:combine}

We describe a method of combining the OVRO mm-array and Nobeyama 45\,m 
telescope data to obtain high-spatial-resolution and missing-flux free images.
We used the 45\,m data to fill the central \uv\ hole of the mm-array
visibility data. 
Since the OVRO array consists of six 10.4\,m diameter telescopes,
we can attain the minimum projected baseline length normalized 
by the observing wavelength ($D_{\lambda\rm ,int}^{\rm min}$) of 
$\simeq 3.5 ~k\lambda$ at 3\,mm 
when the array takes the most compact ``C'' configuration.
The minimum spatial frequency is sufficiently smaller than the maximum 
spatial frequency of \Dlambdanyquist$= 15 ~k\lambda$ for a spatially Nyquist sampled 
45\,m telescope map.
Therefore, the usage of the 45\,m maps works to recover
the flux densities missed by the mm-array.
The basics of the data combination are described in Vogel et al. (1984), 
and our method here is similar to those in the
previous works (e.g., Takakuwa et al. 2003).
We took the following steps to make the combined maps.

\begin{enumerate}

\item\label{itm:grid} For the BEARS \HtCOp\ data, we used the spectra taken with the inner 9 beams 
centered on the central beam on axis. The alignment of the 9 beams 
can be approximated as a regular grid with 41\farcs4 spacing
for our RMS pointing accuracy of $\lesssim 2\arcsec$.
This makes the resultant map grid spacing of 20\farcs7, i.e., full-beam sampling.
Obtaining the data on a regular grid is required to use Fast Fourier Transformation (FFT).
As for the \NtwoH, we used the data on a regular grid of 17\arcsec\ spacing 
(full-beam sampling) with pointing accuracy of $\lesssim 2\arcsec$.
To achieve the excellent pointing accuracy, 
we used the spectra under wind speeds of $\lesssim$ 2 m s$^{-1}$.
We corrected the \Tmb\ scale of the \HtCOp\ data with the BEARS in double sideband mode
by observing the core center with the S80/S100 receivers having single sideband filters.
Since \Tmb\ at the core center is 0.73$\pm$0.04 K 
by the S80/S100 receivers with 18\farcs5 beam
and 0.91$\pm$0.1 K by the BEARS with 20\farcs3 beam,
we scaled the intensities of the BEARS data by a factor of 0.80.

\item To have a common frequency resolution between the OVRO and 45\,m telescope data,
the 45\,m \NtwoH\ spectra, which have a 37 kHz spectral resolution,
are smoothed in the same frequency bins as those of the OVRO data
with a 125 kHz resolution.
For the \HtCOp\ line, the 45\,m and OVRO data have the same frequency bins.

\item We made the 45\,m maps by full-beam 
sampling on the regular grids with spacings of 20\farcs7 ($=41\farcs4/2$) 
for \HtCOp\ and 17\farcs0 for \NtwoH.
We did not apply the task BPOSC in the NEWSTAR package, 
which corrects the misalignment between the 25 beams on the sky plane, to the \HtCOp\ data.
Note that our OVRO field of view (FoV) with a single pointing does not cover the whole 
extent of the core.

\item In converting the \Tmb\ scale 
of the 45\,m maps into the flux density per beam solid angle (\Fnu),
we calculated conversion factors
through a relation of 
$F_{\nu}=I_\nu\Omega_{\rm beam}$ with the Rayleigh-Jeans approximation.
Here $I_{\nu}$ is the specific intensity and 
$\Omega_{\rm beam}$ is the beam solid angle.
The conversion factors are given by
$\frac{F_{\nu}}{T_{\rm mb}} =  \frac{2k}{\lambda^2}\cdot 
\frac{\pi \theta_{\rm HPBW}^2}{4\ln 2} =  3.27
\left(\frac{3~{\rm mm}}{\lambda}\right )^2
\left(\frac{\theta_{\rm HPBW}}{20\arcsec}\right )^2~({\rm Jy~beam^{-1}~K^{-1}})$.

\item\label{itm:45beam} To deconvolve the 45\,m maps with its beam,
the beam pattern is approximated by an axisymmetric Gaussian profile with 
HPBW of 20\farcs3 at 86.754 GHz for the BEARS \HtCOp\ data and 
17\farcs2 at 93.173 GHz for the S80/S100 \NtwoH\ data.
We adopt a mean HPBW of the inner 9 beams for BEARS ($20\farcs3\pm 0\farcs3$) \HtCOp\ observations.
Note that the S80/S100 receivers have nearly equal beam pattern as
they detect two independent linearly polarized waves by sharing common dewar and waveguide.

\item\label{itm:SDdeconvolve} The 45\,m maps are deconvolved 
by the Gaussian beams defined in (\ref{itm:45beam}).
Since our 45\,m maps satisfy the criteria discussed in 
Vogel et al. (1984), the deconvolution is done by dividing the maps by the beam pattern.
After several trials including the steps described below,
the most plausible combined maps are obtained when the
cut-off level of deconvolution is set to be 20\% of the RMS noise level of the single-dish map
so as not to miss high spatial frequency components having low amplitudes.

\item The deconvolved 45\,m maps are convolved with the primary 
beam pattern of the OVRO 10.4\,m
antennas, assuming that all the six element telescopes have 
an axisymmetric Gaussian profile with 
HPBWs of 69\farcs7 at 93.174 GHz and 75\farcs0 at 86.754 GHz.

\item The single-dish maps are converted into visibility data in the
\uv\ domain by FFT. As demonstrated in Figure \ref{afig:uvamp},
the visibility amplitudes from the two instruments 
agree well with each other over the common \uv\ distances.
We stress that we did not apply amplitude scale corrections to both the 45\,m and OVRO data.

\item The FFTed single-dish visibilities are used to make 
model visibilities which have the same integration time as the OVRO data (205 sec).
Since our 45\,m maps are sampled with the full-beam spacing, $\theta^{\rm F}_{\rm SD}$,
we made model visibilities between zero and 
\Dlambdafull\ $= 1/(2\theta^{\rm F}_{\rm SD})$ on the basis of the sampling theorem.
We calculated
\Dlambdafull$=$ 6.10 and 5.08 $k\lambda$ for the \NtwoH\ and \HtCOp\ data, respectively,
and conservatively adopted smaller cut-off values of
5.3 and 4.5 $k\lambda$ to fill the central \uv\ holes in the OVRO data (Table \ref{tbl:combine}).

\item We combined the 45\,m and OVRO visibilities in the \uv\ plane, assuming
that the 45\,m model visibilities have a Gaussian distribution within a radius of 
\Dlambdafull\ in the \uv\ plane.
We found that synthesized beams of the combined maps are controlled
by the visibility weighting function, 
regardless of the distribution of single-dish model visibilities.

\item To construct an image from the combined visibilities, 
we used the task IMAGR in the AIPS package where one can optimize the synthesized beam size and
sensitivity of the resultant image by changing the robustness power of
the weighting function. Figure \ref{afig:robust} represents 
how the parameter affects the synthesized beam pattern and the dirty map
(map affected by the side lobe pattern of the synthesize beam).
Considering the trade-off between the beam size and the smoothness of the map,
we selected the robustness power of $-1$ and CLEANed the dirty maps.
We stopped the CLEAN loop before encountering the first negative CLEAN component
because we are unable to find emission-free regions to estimate an RMS noise level, 
i.e., a threshold flux density to stop the loop, due to the extent of the core 
than the OVRO FoV.

\end{enumerate}

\section{Calculation of Column Density}
\label{as:Ncol}

To estimate the total column density of molecular gas ($N_{\rm tot}$) through
observations of the rotational transition from the upper $J+1$ to lower $J$ level,
we use the previously derived equation 
(e.g., Scoville et al. 1986)
which gives $N_{\rm tot}$ through the optical depth ($\tau_v$) 
and excitation temperature (\Tex) of the probe molecule as follows,

\begin{equation}
N_{\rm tot} = \frac{3k}{8\pi^3\mu^2 B}\cdot\frac{(T_{\rm ex}+\frac{hB}{3k})}{J+1}\cdot
\frac{\exp\left\{\frac{hBJ(J+1)}{kT_{\rm ex}}\right\}}{1-\exp\left(-\frac{h\nu}{kT_{\rm ex}}\right)} 
\int\tau_{v} dv 
\label{eq:Ntot}
\end{equation}

where
$k$ denotes the Boltzmann constant,
$h$ the Planck constant, 
$\mu$ the permanent dipole moment of the molecule, and
$B$ rotational constant,
respectively.
We should keep in mind that, to derive Eq.(\ref{eq:Ntot}), 
one has approximated the rotational partition function
$Q(T_{\rm ex})$ as,

\begin{equation}
Q(T_{\rm ex})\simeq
\frac{k}{hB}
\left\{T_{\rm ex}+\frac{hB}{3k}\right\}.
\label{eq:Qrot}
\end{equation}

\subsection{\NtwoH }
\label{ass:NtwoH}

To apply Eq.(\ref{eq:Ntot}) to the \NtwoH\ line data, 
we have to find \Tex\ and $\tau_{\nu}$.
In the following, 
we describe how they are obtained from the \NtwoH\ spectra through the 
hyperfine structure (HFS) analysis.
In local thermodynamic equilibrium (LTE), 
an observed brightness temperature (\Tb) is expressed by
the following radiative transfer equation,

\begin{equation}
T_{\rm b} = f\cdot[J_{\nu}(T_{\rm ex}) -J_{\nu}(T_{\rm bg})]\{1-e^{-\tau_\nu}\} ,
\label{eq:LTE}
\end{equation}

where \Tb\ is replaced by
the brightness temperatures averaged over a main beam (\Tmb)
for single-dish observations, and over a synthesized beam (\Tsb) 
for interferometric observations.
Here $f$ denotes a beam filling factor, and
$J_{\nu}(T)$ the equivalent Rayleigh-Jeans temperature given by
$J_{\nu}(T) = \frac{h\nu}{k}
\left \{ 
    \exp\left (\frac{h\nu}{kT}\right )-1
\right \}^{-1}$.
From Eq.(\ref{eq:LTE}), we fit the function of 
$y(v)=C\left [1-\exp\left \{-\tau (v)\right \}\right ]$ with a constant $C$
to each \NtwoH\ hyperfine (HF) component.
The optical depth is given by,

\begin{equation}
\tau(v) = 
\tau_{\rm tot}\mathop\sum_{i=1}^{7}s_i
\exp\left [ -
\frac{\left \{v-(V_{\rm sys}+\delta v_i)\right \}^2}{2\sigma_i^2} 
\right ] , \\
\label{eq:tauv}
\end{equation}

as a function of velocity. Here \taut\ denotes the total optical depth, i.e., 
a sum of the optical depths at the centers of the HF components,
$s_i$ the normalized relative intensity of the $i$th HF component
taken from Cazzoli et al. (1985), 
$\delta v_i$ the velocity offset of the $i$th component, 
and $\sigma_i^2 = \frac{(\Delta v_{\rm FWHM,{\it i}})^2}{8\ln 2}$
with the velocity width in FWHM of the $i$th component ($\Delta v_{\rm FWHM,\it i}$).
We assumed that the line broadening due to the thermal and non-thermal gas motions 
equally work for all the HF components since they are emanated from the same molecular ion
in the same region.
We thus simplified Eq. (\ref{eq:tauv}) by adopting
$\sigma_i=\sigma$ and 
$\Delta v_{\rm FWHM,{\it i}}=\Delta v_{\rm FWHM}$.
Therefore, the HFS analysis gives $\tau_{\rm tot}$, 
the center velocity $(V_{\rm sys}+\delta v_i)$, 
the velocity width $(\Delta v_{\rm FWHM})$, and the constant $C$.
Since $C = f\{\JTex\ -\JTbg\ \}$, one can calculate \Tex\ by,

\begin{equation}
T_{\rm ex}=\frac{h\nu_0}{k}
\left[\ln\left\{
1+\frac{h\nu_0}{k\left (C+J_{\nu}\left (T_{\rm bg}\right )\right )}
\right\}\right ]^{-1} .
\label{eq:Tex}
\end{equation}

Here we assumed $f = 1.0$
because the region of our interest is larger than the beam sizes.
We now have the estimates of \Tex, \taut, and \dVint\ at each observing point
(each pixel for the combined data).
Since the integral in Eq.(\ref{eq:Ntot}) becomes
\begin{equation}
\int \tau_vdv =
\frac{1}{2}\sqrt{\frac{\pi}{\ln 2}}
\mathop\sum_{i=1}^{7} \tau_i \Delta v_{\rm int} =
\frac{1}{2}\sqrt{\frac{\pi}{\ln 2}}\cdot \tau_{\rm tot}\Delta v_{\rm int} ,
\label{eq:IntdV}
\end{equation}
we have the desired expression to calculate the column density of 
\NtwoH\ (\NNtwoH) as follows,

\begin{equation}
N_{\rm N_2H^+} = 3.30\times 10^{11}
\frac{(T_{\rm ex}+0.75)}{1-e^{-4.47/T_{\rm ex}}}
\left( \frac{\tau_{\rm tot}}{1.0}\right )
\left( \frac{\Delta v_{\rm int}}{1.0~\rm km~s^{-1}}\right )
~({\rm cm^{-2}}) .
\label{eq:NNtwoHfinal}
\end{equation}

Here we used
$\mu=3.4$ D (Havenith et al. 1990) and
$B=46586.702$ MHz (Caselli, Myers \& Thaddeus 1995).

\subsection{\HtCOp}
\label{ass:HtCOp}

We estimated the column density of \HtCOp\ (\NHtCOp) with Eq.(\ref{eq:Ntot}) giving 
the mean \Tex\ and $\tau_{\nu_0}$ over the full extent of the core
(Table \ref{tbl:ResultsLine}) where $\tau_{\nu_0}$ is a mean optical depth at the line center frequency of $\nu_0$.
Since the radiative transfer equation of Eq.(\ref{eq:LTE}) is written as
$T_{\rm b} = 
f\cdot[J_{\nu}(T_{\rm ex}) -J_{\nu}(T_{\rm bg})]\cdot\tau_\nu\frac{(1-e^{-\tau_\nu})}{\tau_\nu}$, 
the integral of $\tau_\nu$ can be expressed as, 

\begin{equation}
\int \tau_\nu \left (v\right ) dv \simeq
\frac{\tau_{\nu_0}}{1-e^{-\tau_{\nu_0}}}\cdot\frac{\int T_{\rm b}\left (v\right ) dv}{J_{\nu}(T_{\rm ex}) -J_{\nu}(T_{\rm bg})} {\rm ~~~for~~\tau_\nu \simeq 0} .
\label{eq:tau_approx}
\end{equation}

Recall that \HtCOp\ (1--0) in \gf\ is optically thin ($\S\ref{ss:col_h13co}$) and 
we assumed $f=1$ in $\S$\ref{ass:NtwoH}.
Combining Eqs.(\ref{eq:Ntot}) and (\ref{eq:tau_approx}), 
we obtain an expression to estimate the \HtCOp\ column density from single-dish data as,

\begin{equation}
N_{\rm H^{13}CO^+} = 2.32\times 10^{11} 
\frac{(T_{\rm ex}+0.69)}{1-e^{-4.16/T_{\rm ex}}}\cdot 
\frac{1}{J_{\nu}(T_{\rm ex}) -J_{\nu}(T_{\rm bg})}\cdot
\left(\frac{\int T_{\rm mb} dv}{1.0 {\rm ~K~ km s^{-1}}}\right) 
~({\rm cm^{-2}}) .
\label{eq:NHtCOpTmb}
\end{equation}

Here we took $\mu=4.07$ D (Haese \& Woods 1979) and $B=43377.17$ MHz.
To have a complementary expression for interferometric data, 
we converted the flux density ($F_{\nu}$) per synthesized beam into 
a brightness temperature per beam (\Tsb) through
$T_{\rm sb}=\frac{\lambda^2}{2k\Omega_{\rm b}}\cdot F_{\nu}$
where \Omegabeam\ is a beam solid angle given by
$\frac{\pi}{4\ln 2}\left( \theta_{\rm maj}\times\theta_{\rm min}\right)$.
Therefore, we have 

\begin{equation}
N_{\rm H^{13}CO^+} = 3.77\times 10^{13} 
\frac{(T_{\rm ex}+0.69)}{1-e^{-4.16/T_{\rm ex}}}\cdot 
\frac{1}{J_{\nu}(T_{\rm ex}) -J_{\nu}(T_{\rm bg})}\cdot
\left(\frac{\Omega_{\rm b}}{1\arcsec\times 1\arcsec}\right)^{-1}
\left(\frac{\int F_{\nu} dv}{1.0 {\rm ~Jy~ km s^{-1}}}\right)
~({\rm cm^{-2}}) .
\label{eq:NHtCOpSnu}
\end{equation}

\subsection{CCS}
\label{ass:CCS}

Column density analysis of the CCS $4_3-3_2$ line is similar with that 
for the \HtCOp\ line in the sense that
we have only the representative values of \Tex\ and $\tau_{\nu_0}$ 
over the entire core (Table \ref{tbl:ResultsLine}).
The observed CCS brightness temperature are converted into the column density through

\begin{equation}
N_{\rm CCS} = 8.16\times 10^{11} 
\frac{(T_{\rm ex}+0.73) \exp\left (\frac{3.73}{T_{\rm ex}}\right) }{1-e^{-2.18/T_{\rm ex}}} \cdot 
\frac{\tau_{\nu_0}}{1-e^{-\tau_{\nu_0}}}\cdot 
\frac{1}{J_{\nu}(T_{\rm ex}) -J_{\nu}(T_{\rm bg})}\cdot
\left(\frac{\int T_{\rm mb} dv}{1.0 {\rm ~K~ km s^{-1}}}\right)
~({\rm cm^{-2}}) .
\label{eq:Nccs}
\end{equation}

Here we used $\mu=2.81$ D (Murakami 1990), and $B=6477.75036$ MHz (Yamamoto et al. 1990).
Although the rotational energy (\Er) of the CCS molecule is given by 
$E_{\rm r}=hBJ(J+1)-hDJ^2(J+1)^2$ considering the centrifugal stretching
with $D=1727.96$ kHz (Yamamoto et al. 1990),
we ignored the $D$ term to use the rotational partition function of Eq.(\ref{eq:Qrot}).

\clearpage
\begin{deluxetable}{cccccccccc}
\tablewidth{0pt}
\tabletypesize{\scriptsize}
\tablecaption{Summary of Single-Dish Telescope Observations
\label{tbl:sdobs}}
\tablehead{
\colhead{\lw{Emission}}         & \colhead{Rest Frequency} & 
\colhead{\lw{Telescope}}        & \colhead{\lw{Receiver}} & 
\colhead{$\theta_{\rm HPBW}$\tablenotemark{a}}   & \colhead{\lw{$\eta_{\rm mb}$\tablenotemark{b}}} & 
\colhead{$\Delta v_{\rm res}$\tablenotemark{c}}  & \colhead{$\sigma_{\rm T_{\rm A}^*}$\tablenotemark{d}} & 
\colhead{Map\tablenotemark{e}}  & \colhead{Area\tablenotemark{f}} \\
                                & \colhead{(MHz)}     &           
                                &           & 
\colhead{(arcsec)}              &           & 
\colhead{(km s$^{-1}$)}         & \colhead{(mK)}    &    
\colhead{Type}                  &  
\colhead{(arcmin)}                      \\
}
\startdata
NH$_3$ (1,1)               & 23694.495  & NRO  & H22      & 78   & 0.82  & 0.477  &  46 & M/N  & $5.9\times 5.9$ \\
NH$_3$ (2,2)               & 23722.633  & NRO  & H22      & 78   & 0.82  & 0.476  &  56 & M/N  & $5.9\times 5.9$ \\
NH$_3$ (3,3)               & 23870.130  & NRO  & H22      & 78   & 0.82  & 0.473  &  58 & M/N  & $5.9\times 5.9$ \\
NH$_3$ (4,4)               & 24139.417  & NRO  & H22      & 78   & 0.82  & 0.469  &  62 & M/N  & $5.9\times 5.9$ \\
CC$^{34}$S $4_3-3_2$       & 44497.559  & NRO  & S40      & 40   & 0.76  & 0.253  &  25 & D    & \nodata  \\
CCS $4_3-3_2$              & 45379.033  & NRO  & S40      & 40   & 0.76  & 0.245  &  50 & M/F  & $5.3\times 5.3$ \\
HC$^{18}$O$^+$ (1--0)      & 85162.157  & NRO  & S80      & 18.7 & 0.50  & 0.130  &  26 & D    & \nodata  \\
C$_3$H$_2$ $2_{12}-1_{01}$ & 85338.906  & NRO  & S100     & 18.7 & 0.50  & 0.130  &  30 & D    & \nodata  \\
H$^{13}$CO$^+$ (1--0)      & 86754.330  & NRO  & S80/S100 & 18.5 & 0.50  & 0.128  &  37 & C/F  & $3.7\times 2.4$  \\
                           & 86754.330  & NRO  & BEARS    & 20.3 & 0.49  & 0.108  &  48 & M/F  & $8.6\times 8.6$  \\
SiO (1--0) $v=0$           & 86846.998  & NRO  & S80      & 18.5 & 0.50  & 0.128  &  42 & M/F  & $3.7\times 2.4$   \\
                           & 86846.998  & NRO  & BEARS    & 20.5 & 0.49  & 0.108  & 185 & M/F  & $8.6\times 8.6$   \\
N$_2$H$^{+}$(1--0)         & 93173.900  & NRO  & S80/S100 & 17.2 & 0.51  & 0.119  &  31 & M/F  & $2.0\times 2.0$   \\
$^{12}$CO (1--0)           & 115271.204 & NRO  & S100     & 14.8 & 0.49  & 0.0976 &  94 & C/F  & $2.8\times 1.8$   \\
                           & 115271.204 & NRO  & BEARS    & 16.5 & 0.51  & 0.0983 & 139 & M/F  & $7.1\times 7.1$   \\
$^{12}$CO (3--2)           & 345795.989 & CSO  & \nodata\ & 22.0 & 0.75  & 0.416  & 290 & M/F  & $3.7\times 3.0$   \\
\um\ continuum             & 850000     & CSO  & \sharc\  & 8.9$\times$8.0 & \nodata & \nodata & \nodata & \nodata & $4.0\times 3.8$   \\
\enddata
\tablenotetext{a}{Half-power beam width for a circular Gaussian beam}
\tablenotetext{b}{Main beam efficiency}
\tablenotetext{c}{Effective velocity resolution}
\tablenotetext{d}{Typical RMS noise level of the spectrum. 
For detection observations, an RMS noise level of the spectrum is given. 
For mapping observations, a mean RMS noise level for all the spectra is given.}
\tablenotetext{e}{Observing modes for the molecular line emission; 
D denotes ``detection'' observations toward the center position of 
R.A.(J2000)$=20^h51^m29.5^s$, Decl.(J2000)$=60^d18'38.0''$, 
M/F and M/N denote, respectively, ``mapping'' observations with full-beam and 
Nyquist sampled grid spacings centered on the above position, 
and C/F denotes ``cross scan'' observations with 
a full-beam spacing along the NE--SW and NW--SE directions centered on the position.}
\tablenotetext{f}{Size of the region for the ``mapping'' observations, 
and the major and minor axis lengths for ``cross scan'' ones.}
\end{deluxetable}

\begin{deluxetable}{ccccrccc}
\tablewidth{0pt}
\tabletypesize{\scriptsize}
\tablecaption{Summary of Interferometric Observations
\label{tbl:cont_obs}}
\tablewidth{0pt}
\tablehead{
                  & \colhead{} & 
\colhead{Center}  & 
\multicolumn{2}{c}{Synthesized Beam} & 
\colhead{\lw{Bandwidth}} &
\colhead{Velocity} &
\colhead{Image} \\
\cline{4-5}
\colhead{Array}              & \colhead{Emission} & 
\colhead{Frequency}          & \colhead{$\theta_{\rm maj}\times\theta_{\rm min}$} &
\colhead{P.A.}               & \colhead{} &
\colhead{Resolution} & 
\colhead{Sensitivity} \\
                             & \colhead{} & 
\colhead{(MHz)}              & \colhead{(arcsec$\times$arcsec)} & 
\colhead{(deg)}              & \colhead{(MHz)} &
\colhead{(\kms)}             & \colhead{(mJy \pbeam)} \\
}
\startdata
VLA  & continuum      &  8435    & 3.29$\times$2.93 &   29.7  &  172\tablenotemark{a} & \nodata & 0.038 \\
OVRO & continuum      & 89964    & 5.38$\times$4.80 &   41.3  & 4096\tablenotemark{b} & \nodata & 0.15 \\
     & \NtwoH\ (1--0) & 93173.78 & 4.91$\times$4.36 & $-62.4$ & 7.75                  & 0.402   & 17.6\tablenotemark{c} \\
     & \HtCOp\ (1--0) & 86754.33 & 6.31$\times$4.84 & $-24.7$ & 1.938                 & 0.108   & 33.0\tablenotemark{c} \\
\enddata
\tablenotetext{a}{Concatenated the two polarization data.}
\tablenotetext{b}{Concatenated the lower and upper sideband data.}
\tablenotetext{c}{Per velocity channel.}
\end{deluxetable}

\begin{deluxetable}{lccccccc}
\tablewidth{0pt}
\tabletypesize{\scriptsize}
\tablecaption{Summary of the Combined Data from the OVRO MM-Array and the NRO 45 M Telescope\tablenotemark{a}
\label{tbl:combine}}
\tablewidth{0pt}
\tablehead{ 
\colhead{} & 
\multicolumn{2}{c}{Spatial Frequency Range} & &
\colhead{Field of} &   
\multicolumn{2}{c}{Synthesized Beam} & 
\colhead{Velocity} \\
\cline{2-3}
\cline{6-7}
\colhead{Line} & 
\colhead{45\,m } & \colhead{OVRO}  & &
\colhead{View} & 
\colhead{$\theta_{\rm maj}\times\theta_{\rm min}$} &                  
\colhead{P.A.} & 
\colhead{Resolution} \\
\colhead{} &
\colhead{(k$\lambda$)} & 
\colhead{(k$\lambda$)} & 
\colhead{} &
\colhead{(arcsec)} & 
\colhead{(arcsec)} & 
\colhead{(deg)} & 
\colhead{(km s$^{-1}$)} \\
}
\startdata
N$_2$H$^{+}$(1--0)    & 0.0--5.29 & 3.30--71.8 & & 69.7 & 4.49$\times$3.86 & $-70.7$ & 0.402 \\
H$^{13}$CO$^+$ (1--0) & 0.0--4.48 & 3.07--58.0 & & 75.0 & 6.73$\times$5.14 & $-27.2$ & 0.108 \\
\enddata
\tablenotetext{a}{For details, see Appendix \ref{as:combine}}
\end{deluxetable}

\begin{deluxetable}{rccccc}
\tablewidth{0pt}
\tabletypesize{\scriptsize}
\tablecaption{Results of Continuum Emission Flux Measurements
\label{tbl:ResCont}}
\tablewidth{0pt}
\tablehead{
\colhead{\lw{$\lambda$}} & \colhead{\lw{Instruments}} & 
\multicolumn{2}{c}{Size} & \colhead{Peak} & \colhead{Total Flux} \\
\cline{3-4}
                    &               & \colhead{$l_{\rm maj}\times l_{\rm min}$} & \colhead{P.A.} & \colhead{Intensity} & \colhead{Density} \\
\colhead{($\mu$m)}  &               & \colhead{(arcsec$\times$arcsec)} & \colhead{(deg)} & \colhead{(mJy beam$^{-1}$)} & \colhead{(mJy)} 
}
\startdata
35540    & VLA-D       & \nodata                             & \nodata      &       $ <0.055$ & \nodata  \\
 3330    & OVRO        & $(6.1\pm 0.9)\times (4.3\pm 0.6)$   & 129$\pm$17 &   1.16$\pm$0.17 & 0.75$\pm$0.07 \\
  350    & CSO-SHARCII & $(32.9\pm 4.3)\times (22.5\pm 2.9)$ & 140$\pm$13 &   300$\pm$30     & 1830$\pm$330   \\
\enddata
\end{deluxetable}

\begin{deluxetable}{lllclccl}
\tablewidth{0pt}
\tabletypesize{\scriptsize}
\tablecaption{Results from the 45\,m Telescope Spectral Line Observations: Dense Gas Tracers
\label{tbl:ResultsLine}}
\tablewidth{0pt}
\tablehead{
\colhead{}               & 
\multicolumn{3}{c}{Core Center\tablenotemark{a}} & &
\multicolumn{3}{c}{Inside the Half-Maximum Contour\tablenotemark{b}} \\
\cline{2-4}\cline{6-8}
\colhead{Line}                & 
\colhead{$T_{\rm mb}$\tablenotemark{c}} & 
\colhead{$\Delta v_{\rm int}$\tablenotemark{d}} & 
\colhead{\lw{$\tau$\tablenotemark{e}}} & 
                                  &
\colhead{$\langle \Delta v_{\rm int}\rangle$\tablenotemark{d}} &
\colhead{\lw{$\langle \tau\rangle$\tablenotemark{e}}} &
\colhead{$\langle T_{\rm ex}\rangle$} \\
                              & 
\colhead{(K)}                 &  
\colhead{(km s$^{-1}$)}       &     
                              & 
                              &
\colhead{(km s$^{-1}$)}       &
                              &
\colhead{(K)}                 \\
}
\startdata
N$_2$H$^{+}$(1--0)         & 1.96$\pm$0.05 & 0.35$\pm$0.01 & 3.4$\pm$0.08  & & 0.32$\pm$0.09  & 2.7$\pm$0.4   & 9.5$\pm$1.9 \\
H$^{13}$CO$^+$ (1--0)      & 0.88$\pm$0.09 & 0.60$\pm$0.03 & $\leq 0.1$    & & 0.60$\pm$ 0.17 & \nodata       & \nodata \\
C$_3$H$_2$ $2_{12}-1_{01}$ & 1.12$\pm$0.06 & 0.59$\pm$0.1  & 0.34          & & \nodata        & \nodata       & \nodata \\
CCS $4_3-3_2$              & 0.74$\pm$0.07 & 0.41$\pm$0.04 & 0.94          & & 0.41$\pm$0.19  & \nodata       & \nodata \\ 
NH$_3$ (1,1)               & 1.20$\pm$0.06 & 0.61$\pm$0.10 & 1.6$\pm$0.1   & & 0.62$\pm$0.17  & 0.45$\pm$0.18 & $\leq 9.7$ \\
\enddata
\tablenotetext{a}{Single pointing toward the 3\,mm continuum source. 
See the spectra in Figure \ref{fig:spectra}, and beam sizes in Table \ref{tbl:sdobs}.}
\tablenotetext{b}{For \NtwoH\ and \amm\ lines which allow to derive optical depth ($\tau$) and
excitation temperature (\Tex) through hyperfine structure (HFS) analysis, 
we present mean and standard deviation for the spectra existing inside the half-maximum, i.e.,
50\% level contour of total integrated intensity map 
(Figure \ref{fig:tiimaps}).}
\tablenotetext{c}{Peak main-beam brightness temperature.
For \NtwoH\ and \amm\ lines, those for the brightest HF components are given.}
\tablenotetext{d}{In calculating intrinsic velocity width (\dVint), we employed
HFS analysis for the \NtwoH\ and \amm\ lines,
spectral moment analysis for the \HtCOp\ and \CtHt\ lines, 
and single Gaussian fitting for the CCS line ($\S$\ref{ss:widthmap}).
For the \CtHt\ and CCS lines, we corrected for line broadening due to the
optical depth (Phillips et al. 1969).
}
\tablenotetext{e}{Optical depth. We present $\tau$ of
the brightest hyperfine transition \tbrightest\ for \NtwoH\ (1--0) ($\S\ref{ss:col_n2h}$) and
F,F$_1$$=\frac{5}{2},2-\frac{5}{2},2$ for \amm\ (1,1) ($\S\ref{ss:col_amm}$).}
\end{deluxetable}

\begin{deluxetable}{lccccccc}
\tablewidth{0pt}
\tabletypesize{\scriptsize}
\tablecaption{Column Densities of \NtwoH, \HtCOp, \CtHt, CCS, and \amm\ Molecules\tablenotemark{a}
\label{tbl:Ncol}}
\tablewidth{0pt}
\tablehead{
\colhead{} & 
\colhead{Combined} & 
\colhead{} &         
\multicolumn{5}{c}{Nobeyama 45\,m Telescope} \\
\cline{2-2}
\cline{4-8}
\colhead{} & 
\colhead{\lw{Inside \um}} & 
\colhead{} &
\multicolumn{3}{c}{\lw{Core Center\tablenotemark{c}}} &
\colhead{} & 
\colhead{\lw{Full Extent of}} \\
\multicolumn{1}{l}{Line} & 
\colhead{\lw{3$\sigma$ Contour\tablenotemark{b}}} &  
\colhead{} &
\multicolumn{3}{c}{} &
\colhead{} & 
\colhead{\lw{the Core\tablenotemark{d}}} \\
\cline{4-6}
\colhead{} & 
\colhead{} & 
\colhead{} &
\colhead{\lw{$r\leq 10\arcsec$}}         & 
\colhead{\lw{$r\leq 20\arcsec$}}         & 
\colhead{\lw{$r\leq 40\arcsec$}}         &
\colhead{} & 
\colhead{} \\
\colhead{} & 
\colhead{\lw{(\cmq)}} & 
\colhead{} &
\colhead{\lw{(\cmq)}} &
\colhead{\lw{(\cmq)}} &
\colhead{\lw{(\cmq)}} &
\colhead{} & 
\colhead{\lw{(\cmq)}}             \\
}
\startdata
N$_2$H$^{+}$    & (5.2$\pm$2.5)E+13  & & (5.6$\pm$0.2)E+13 & (4.4$\pm$0.3)E+13 & (3.5$\pm$0.2)E+13 & & (3.9$\pm$1.0)E+13 \\
H$^{13}$CO$^+$  & (4.7$\pm$0.8)E+11  & & (6.1$\pm$0.3)E+11 & \nodata           & \nodata           & & (4.3$\pm$0.2)E+11 \\
C$_3$H$_2$      & \nodata            & & (2.4$\pm$0.3)E+13 & \nodata           & \nodata           & & \nodata  \\
CCS             & \nodata            & & \nodata           & (1.5$\pm$0.2)E+13 & \nodata           & & (1.5$\pm$0.3)E+13 \\
NH$_3$          & \nodata            & & \nodata           & \nodata           & (8.2$\pm$0.2)E+14 & & (7.7$\pm$2.8)E+14 \\
\enddata
\tablenotetext{a}{Calculated with the assumption of \Tex\ $=$ 9.5 K ($\S$\ref{ss:col_n2h})}
\tablenotetext{b}{Column densities measured inside an area enclosed by the 3$\sigma$ level 
contour of the \um\ continuum emission.
For this purpose, we made the \NtwoH\ and \HtCOp\ maps by giving the \sharc\ beam size
as a synthesized beam size.}
\tablenotetext{c}{Toward the 3\,mm continuum source measured with the beam size in Table \ref{tbl:sdobs}.
e.g., \thetahpbw\ $=$ 20\arcsec\ beam for the region of $r\leq$ 10\arcsec.
Regarding the \NtwoH\ column densities for the regions of $r\leq$ 20\arcsec\ and 40\arcsec, we calculate
column densities using these spectra smoothed with \thetahpbw\ $=$ 40\arcsec\ and 80\arcsec\ beams, respectively 
($\S\ref{ss:col_n2h}$).} 
\tablenotetext{d}{Measured over the full extent of the core with 
beam sizes of 20\arcsec\ for \NtwoH, \HtCOp, and \CtHt, 40\arcsec\ for CCS, and 80\arcsec\ for \amm.
Here the full extent of the core is defined by a region enclosed by the 50\% level contour of
each total intensity map.}
\end{deluxetable}

\begin{deluxetable}{clcl}
\tablewidth{0pt}
\tabletypesize{\scriptsize}
\tablecaption{Fractional Abundances of Dense Gas Tracer Molecules
\label{tbl:Xabs}}
\tablewidth{0pt}
\tablehead{
\colhead{\lw{Molecule}} & 
\colhead{\lw{X$_{\rm m}$\tablenotemark{a}}} &
\colhead{Beam Size} &
\colhead{\lw{Comments}} \\
\colhead{} & 
\colhead{} &
\colhead{(arcsec)} &
\colhead{} \\
}
\startdata
\NtwoH\         & \XNtwoHEE\ & $9\times 8$ & Used \meanNHmol\tablenotemark{b} in $\S\ref{ss:rescont_dust}$ \\
\HtCOp\         & \XHtCOpEE\ & $9\times 8$ & Used \meanNHmol\tablenotemark{b} in $\S\ref{ss:rescont_dust}$ \\
\CtHt\          & \XCtHtEE\  & 20              & X(N$_2$H$^+)_{8\arcsec}\times \frac{N\rm [C_2H_2]}{N\rm [N_2H^+]} (20\arcsec )$\tablenotemark{c} \\
CCS             & \XccsEE\   & 40              & X(N$_2$H$^+)_{8\arcsec}\times \frac{N\rm [CCS]}{N\rm [N_2H^+]} (40\arcsec )$\tablenotemark{c} \\
\amm\           & \XammEE\   & 80              & X(N$_2$H$^+)_{8\arcsec}\times \frac{N\rm [\rm NH_3]}{N\rm [N_2H^+]} (80\arcsec )$\tablenotemark{c} \\
\enddata
\tablenotetext{a}{Fractional abundance of the molecule.}
\tablenotetext{b}{We adopted mean H$_2$ column density, \meanNHmol, from 
cases I and II in $\S\ref{ss:rescont_dust}$
and considered the range of H$_2$ column densities into the errors in X$_{\rm m}$.}
\tablenotetext{c}{Assumed that \NtwoH\ does not have radial abundance variation.
Column column density ratios are calculated from Table \ref{tbl:Ncol}.
The notation of X(N$_2$H$^+)_{8\arcsec}$ indicates \NtwoH\ abundance measured with the \sharc\ beam.}
\end{deluxetable}

\begin{deluxetable}{cclcccc}
\tablewidth{0pt}
\tabletypesize{\scriptsize}
\tablecaption{LTE and Virial Masses of \gf\ Core
\label{tbl:Mcore}}
\tablewidth{0pt}
\tablehead{
\colhead{} & 
\colhead{\lw{${\cal A}$\tablenotemark{a}}} &
\colhead{\lw{$M_{\rm LTE}$\tablenotemark{b}}} &
\colhead{\lw{$R_{\rm eff}$\tablenotemark{c}}} &
\colhead{} & 
\multicolumn{2}{c}{$M_{\rm vir}$\tablenotemark{d}} \\
\cline{6-7}
\colhead{Line} & 
\colhead{} &
\colhead{} &
\colhead{} &
\colhead{} &
\colhead{Case \sc{I}\tablenotemark{e}} &
\colhead{Case \sc{II}\tablenotemark{e}} \\
\colhead{}                    &            
\colhead{(pc$^{2}$)}          &
\colhead{($\Msun$)}           &
\colhead{(pc)}                &
\colhead{}                    & 
\colhead{($\Msun$)}           &
\colhead{($\Msun$)}  \\
}
\startdata
N$_2$H$^{+}$(1--0)    & 2.632E--3 & 1.2$\pm$0.7         & 0.0289 & & 0.39$\pm$0.3 & 0.10$\pm$0.08 \\
H$^{13}$CO$^+$ (1--0) & 3.963E--3 & 1.3$\pm$0.3         & 0.0354 & & 2.9$\pm$1.7  & 0.12$\pm$0.07 \\ 
CCS $4_3-3_2$         & 5.897E--3 & 3.1$\pm$1.9         & 0.0432 & & 1.7$\pm$1.4  & 0.078$\pm$0.06 \\
NH$_3$ (1,1)          & 5.544E--3 & 4.5$\pm$2.4         & 0.0419 & & 2.3$\pm$1.6  & 0.25$\pm$0.15 \\
\enddata
\tablenotetext{a}{Area enclosed by the 50\% level contour of each total integrated intensity map 
(Figure \ref{fig:tiimaps}).}
\tablenotetext{b}{Errors in LTE-mass include those in the column density, area size, and the abundance.}
\tablenotetext{c}{Effective radius defined by $R_{\rm eff}=\sqrt{{\cal A}/\pi}$.}
\tablenotetext{d}{Errors in virial mass are calculated from 
$\left(\delta M_{\rm vir}\right )^2=
\left(\frac{\partial M_{\rm vir}}{\partial \langle\Delta v_{\rm int}\rangle}\right)^2
\left(\delta\langle\Delta v_{\rm int}\rangle\right)^2+
\left(\frac{\partial M_{\rm vir}}{\partial R_{\rm eff}}\right)^2 \left(\delta R_{\rm eff}\right )^2$
where $\delta R_{\rm eff}$ and $\delta\langle\Delta v_{\rm int}\rangle$ are those in 
\Reff\ and \meandVint, respectively. 
The intrinsic velocity width (\dVint) and its error are taken from those for the
entire core in Table \ref{tbl:ResultsLine}.}
\tablenotetext{e}{Cases I and II are, respectively, when 
$\langle \Delta v_{\rm int}\rangle ^2=\langle \Delta v_{\rm thm}\rangle ^2+\langle \Delta v_{\rm turb}\rangle ^2$
and $\langle \Delta v_{\rm int}\rangle ^2=\langle \Delta v_{\rm thm}\rangle ^2$.
Here $\langle \Delta v_{\rm int}\rangle$, $\langle \Delta v_{\rm thm}\rangle$, 
and $\langle \Delta v_{\rm turb}\rangle$
are means of intrinsic, thermal, and turbulent velocity widths, respectively
(see $\S\ref{ss:stab}$).
}
\end{deluxetable}

\clearpage





\begin{figure}
\begin{center}
\end{center}
\caption{Overlay of the 3\,mm continuum emission map observed with the OVRO mm-array
(grey scale plus contours) on the 350 $\mu$m map with the CSO \sharc\ camera (thick contours).
The contours start from the 3$\sigma$ levels with a 3$\sigma$ interval
where $\sigma =0.15$ and 29.7 mJy \pbeam\ for the 3\,mm and \um\ images, respectively.
The dashed contours show the $-3\sigma$ level for the 3\,mm image.
The \um\ image does not have negative contours below the $-2.2\sigma$ level.
The ellipses at the bottom corners indicate the beam sizes.
\label{fig:contmaps}}
\end{figure}

\begin{figure}
\begin{center}
\end{center}
\caption{Spectra of the molecular line emission shown in the main-beam brightness 
temperature (\Tmb) scale toward the 3\,mm source.
The vertical dashed line indicates the systemic velocity (\Vsys ) of
$-2.5$ \kms\ ($\S\ref{ss:resspectra}$).
The horizontal blue and red bars under the \co\ lines indicate the
LSR-velocity ranges for the integrated intensity maps shown in Figure \ref{fig:COBluRed}, 
and the green bars under the \NtwoH\ and \HtCOp\ lines are for the maps
shown in Figure \ref{fig:hikaku}. 
The \NtwoH\ emission seen in 4.5 $\lesssim$ \Vlsr/\kms\ $\lesssim$ 7.2
is the main hyperfine components (see Figure \ref{fig:NtwoHspectra}).
\label{fig:spectra}
}
\end{figure}

\begin{figure}
\begin{center}
\end{center}
\caption{\NtwoH\ (1--0) spectrum (black histogram) taken
with the 45\,m telescope toward the 3\,mm source.
The green histogram represents the best-fit spectrum from the HF analysis 
($\S\ref{ss:col_n2h}$). We assume that 
the LSR velocity of the isolated F$_1$,F$=$0,1-1,2 transition ($\nu_{\rm rest}= 93176.310$ MHz)
is equal to the \Vsys\ of $-2.5$ \kms.
The lengths of vertical lines above the spectrum indicate the expected relative intensities 
at $\tau\simeq$ 0, plotted at the rest frequencies (Cazolli et al. 1985).
The horizontal bars below the spectrum show the velocity ranges integrated 
to obtain maps in Figure \ref{fig:tiimaps}c 
and those in the first row of Figure \ref{fig:hikaku}.
\label{fig:NtwoHspectra}
}
\end{figure}

\begin{figure}
\begin{center}
\end{center}
\caption{Total integrated intensity maps of the 
(a) \HtCOp\ (1--0), (b) CCS $4_3-3_2$, (c) \NtwoH\ (1--0), and (d) \amm\ (1,1) lines
observed with the 45\,m telescope.
The velocity ranges for the integrations are between
\Vlsr\ $=-2.95$ and $-2.1$ \kms\ for \HtCOp, and
$-3.0$ and $-2.1$ \kms\ for CCS.
For the \NtwoH\ and \amm\ emission, 
all the hyperfine emission are integrated over the velocity ranges shown by 
the horizontal bars in 
Figures \ref{fig:NtwoHspectra} and \ref{fig:ammspectra}, respectively.
Contour intervals are 3$\sigma$, 
starting from the 3$\sigma$ levels.
The $1\sigma$ noise levels are 
33.4, 42.5, 27.0, and 24.8 mK$\cdot$\kms\ for the \HtCOp, CCS, \NtwoH,
and \amm\ maps, respectively.
The central star marks the position of the 3\,mm source.
The hatched circles at the bottom-left corner of each panel show beam size
(see Table \ref{tbl:sdobs}), and the small dots represent the observed grid points.
\label{fig:tiimaps}}
\end{figure}

\begin{figure}
\begin{center}
\end{center}
\caption{Velocity channel maps of the \isolated\ and 
\HtCOp\ (1--0) emission taken with the 45\,m telescope (see Table \ref{tbl:sdobs});
each panel has a size of 225\arcsec $\times$225\arcsec.
The central LSR-velocity for each channel is shown at the upper right corner.
Contour intervals are 2$\sigma$, starting from the 3$\sigma$ levels
of the images where $1\sigma$ noise level is 78 mK in \Tmb\ for the \NtwoH\ and 96 mK for \HtCOp.
The star in each panel is the same as in Figure \ref{fig:tiimaps}.
The systemic velocity is \Vlsr\ $=$ $-2.5$ \kms.
\label{fig:NROchmaps}
}
\end{figure}

\begin{figure}
\begin{center}
\caption{\small Comparison of the total integrated intensity maps of 
all the seven hyperfine (HF) emission of \NtwoH\ (1--0) (first row),  
\isolated\ (second row) emission, and 
\HtCOp\ (1--0) (third row) line taken with the
Nobeyama 45\,m telescope (first column), 
the OVRO mm-array (second column), 
and made from the combined 45\,m plus OVRO visibilities (third column).
The central star and the ellipse at the bottom left corner 
in each panel are the same as in Figure \ref{fig:tiimaps}.
The large circles in the second and third columns indicate the field of view.
The \NtwoH\ total emission maps are integrated over the 3 velocity ranges
shown by the horizontal bars in Figure \ref{fig:NtwoHspectra},
and those for \NtwoH\ \tisolated\ and \HtCOp\ maps by the green ones in Figure \ref{fig:spectra}.
The contour intervals are the same as in Figure \ref{fig:contmaps}.
The 1$\sigma$ rms noise levels are
38.5 mK in \Tmb\ for the 45\,m \NtwoH\ total emission map,
42.5 mK for the 45\,m \NtwoH\ \tisolated\ map,
33.5 mK for the 45\,m \HtCOp\ map,
17.4 mJy \pbeam\ for the OVRO \NtwoH\ total emission image,
22.4 mJy \pbeam\ for the OVRO \NtwoH\ \tisolated\ image,
17.2 mJy \pbeam\ for the OVRO \HtCOp\ image,
62.4 mJy \pbeam\ for the combined \NtwoH\ total emission image, 
95.5 mJy \pbeam\ for the combined \NtwoH\ \tisolated\ image, and
35.3 mJy \pbeam\ for the combined \HtCOp\ image.
The noise levels for the combined data correspond to 
170 mK in \Tsb\ for the \HtCOp, 
510 mK for the \NtwoH\ total, and
780 mK for the \NtwoH\ \tisolated\ emission maps.
\label{fig:hikaku}}
\end{center}
\end{figure}

\begin{figure}
\begin{center}
\end{center}
\caption{
Velocity channel maps of the combined OVRO plus 45\,m telescope images of 
the \isolated\ and \HtCOp\ (1--0) lines (see Table \ref{tbl:combine}).
Contour intervals are the same as in Figure \ref{fig:contmaps}, and 
the RMS noise levels are 32.1 and 52.1 mJy \pbeam\ for the \NtwoH\ and \HtCOp\ lines, respectively.
The size for each channel map is 76\farcs8$\times$76\farcs8.
The systemic velocity is \Vlsr\ $=$ $-2.5$ \kms.
\label{fig:cmbchmaps}
}
\end{figure}

\begin{figure}
\begin{center}
\end{center}
\caption{\small Maps of the blue- (left panels) and redshifted
(right panels) \co\ (1--0) (upper panels) and (3--2) (lower panels) emission. 
The (1--0) emission is integrated over the velocity ranges of
$-4.2\leq$ \Vlsr/\kms\ $\leq -3.4$ for the blue and 
$-1.8\leq$ \Vlsr/\kms\ $\leq -1.0$ for the red.
The (3--2) emission is integrated in the ranges of
$-5.8\leq$ \Vlsr/\kms\ $\leq -3.8$ for the blue and
$-1.4\leq$ \Vlsr/\kms\ $\leq +0.6$ for the red.
These velocity ranges are shown in Figure \ref{fig:spectra}.
Contours with color scale are drawn by the 10\% step up to the 90\% level of the peak value.
The lowest contour for the (1--0) maps is the 10\% level, 
corresponding to the 21$\sigma$ and 8.4$\sigma$ levels for the blue and red, respectively.
In the (3--2) maps, the lowest contour is the 50\% level,
corresponding to the 9.0$\sigma$ and 8.8$\sigma$ levels
for the blue and red, respectively.
The white dashed contours shown in all the panels indicate the 5$\sigma$ level contour
of the \HtCOp\ total intensity map (Figure \ref{fig:tiimaps}a), 
and the inner ones in (c) and (d) indicate the 50\% (corresponding to 16.4$\sigma$) level contour.
The central circles with a radius of 40\arcsec\ in (a) and (b) show the
error circle of the \wat\ maser position (Furuya et al. 2003).
The central yellow star indicates the position of the 3\,mm source.
The other symbols are the same as in Figure \ref{fig:tiimaps}.
\label{fig:COBluRed}
}
\end{figure}

\begin{figure}
\begin{center}
\caption{Column density maps of the \NtwoH\ line
obtained from (a) the 45\,m telescope data and (b) the combined OVRO plus 45\,m data
($\S\ref{ss:rescolumn}$) in color scale.
The color scale bars are the same for both the 45\,m and combined data.
The black contours in (a) and white ones in (b) indicate 
the total integrated intensity
maps of the \NtwoH\ emission (Figure \ref{fig:tiimaps}c) and the
\um\ continuum emission (Figure \ref{fig:contmaps}), respectively.
The white circle in (a) and black one in (b) indicate the OVRO filed of view.
\label{fig:Nmaps}
}
\end{center}
\end{figure}

\begin{figure}
\begin{center}
\end{center}
\caption{\amm\ (1,1) and (2,2) spectra observed 
with the 45\,m telescope toward the 3\,mm continuum emission.
In the lower panel, the green histogram represents the
best-fit \amm\ (1,1) spectrum from the HFS analysis ($\S\ref{ss:col_amm}$).
It is assumed that the (1,1) and (2,2) spectra have their brightest
HF components at \Vlsr\ $=$ $-2.5$ \kms.
The vertical lines above the spectra indicate the expected relative 
intensities at $\tau \simeq$ 0 and LSR-velocities of the HF components.
The horizontal bars below the (1,1) spectrum show the velocity ranges 
to obtain the total intensity map shown in Figure \ref{fig:tiimaps}d.
\label{fig:ammspectra}
}
\end{figure}

\begin{figure}
\begin{center}
\epsscale{0.8}
\epsscale{1.0}
\end{center}
\caption{Radial column density profiles of the \NtwoH\ in \gf\ core 
obtained from the Nobeyama 45\,m telescope (blue) and the combined OVRO plus 45\,m 
data (black) (see text). The upper horizontal axis shows the linear scale. 
The right hand vertical axis represents H$_2$ column density converted with 
X(\NtwoH) $=$ \XNtwoHE\  (Table \ref{tbl:Xabs}).
The solid curves indicate the best-fit power-law profiles ($\S\ref{ss:resdensity}$):
$N_{\rm N_2H^+}(r)=(7.3\pm 1.1)\times 10^{14}\left (r/1\farcs0\right )^{-(1.0\pm 0.1)}$ \cmq\ 
for $r\leq 50\arcsec$ for the combined and 
$(8.5\pm 2.5)\times 10^{14}\left (r/1\farcs0\right )^{-(1.1\pm 0.2)}$ \cmq\ 
for $r\leq 80\arcsec$ for the 45\,m data.
The broken curves indicate the beam patterns:
17\farcs2 in HPBW for the 45\,m telescope, and 4\farcs1 for the combined data.
The vertical dotted lines at $r=29\farcs 9$ (corresponding to 0.029 pc) 
indicates the \Reff\ for the \NtwoH\ core (Table \ref{tbl:Mcore}).
}
\label{fig:radprofile}
\end{figure}

\clearpage
\begin{figure}
\begin{center}
\end{center}
\caption{Position-Velocity (PV) diagrams of the 
\isolated\ (upper panels) and \HtCOp\ (1--0) (lower panels) emission for \gf\
core made from the 45\,m telescope (green contour) and the combined data 
(greyscale with black contour).
The left and right panels, respectively, show the diagrams along the major (P.A.$=45^{\circ}$)
and minor (P.A.$=135^{\circ}$) axes, passing through the continuum source.
The right hand vertical axis in each panel shows distance from the 3\,mm source in pc.
The contour intervals are the same as in Figure \ref{fig:tiimaps}.
The intensity is presented in \Tmb\ scale for the 45\,m and
in \Snu\ for the combined data. 
See the caption of Figure \ref{fig:hikaku} for the RMS noise levels. 
The vertical and horizontal dotted lines indicate \Vsys\ $=-2.5$ \kms\
and the position of the 3\,mm source, respectively.
The horizontal dashed lines in (c) and (d) running at $\pm 37\farcs5$
indicate the field of view of the combined data.
The horizontal and vertical bars labeled with V$_{\rm res}$ and $\theta_{\rm beam}$
indicate the velocity resolution and beam size, respectively.
\label{fig:pv}
}
\end{figure}

\begin{figure}
\begin{center}
\epsscale{.60}
\end{center}
\caption{Intrinsic velocity width (\dVint) maps (color scale) of 
(a) \isolated\ and (b) \HtCOp\ (1--0) emission for the 45\,m telescope data ($\S\ref{ss:widthmap}$).
The presented \dVint\ ranges are 
$0.28 \leq\Delta v_{\rm int}/{\rm km~s^{-1}}\leq 0.61$ for the \NtwoH\ and 
$0.44 \leq\Delta v_{\rm int}/{\rm km~s^{-1}}\leq 0.63$ for the \HtCOp.
Contours in (a) and (b) indicate the \NtwoH\ and \HtCOp\ total integrated intensity
maps shown in Figure \ref{fig:hikaku}.
The thick white contour in each panel indicates the 50\% level contour of
the total maps. The star and hatched ellipse are the same as in Figure \ref{fig:tiimaps}.
\label{fig:dVint}
}
\end{figure}

\begin{figure}
\begin{center}
\epsscale{.60}
\end{center}
\caption{Plot of normalized CO outflow momentum flux 
($F_{\rm CO}\cdot c/L_{\rm bol}$) vs. bolometric temperature ($T_{\rm bol}$)
for nearby ($d<350$ pc) low-mass YSOs less luminous than 
$50 L_{\odot}$ ($\S$\ref{ss:evl_flow}), including S106\,FIR 
($L_{\rm bol}\lesssim 1080 L_{\sun}$; Richer et al. 1993; Furuya et al. 2000).
The filled- and open circles represent the presence and the absence of 
\wat\ masers in the period of the Nobeyama survey (Furuya et al. 2001; 2003), respectively.
The locus of \gf\ ($L_{\rm bol}\simeq 0.3$ \Lsun) is indicated by large filled circle.
The upper limit in \FcocLbol\ for S106\,FIR is recalculated from Furuya et al. (2000)
with the $3\sigma$ upper limit and the outflow velocity of 50 \kms.
The \Tbol\ is taken from Chen et al. (1995) and AWB00, and \FcocLbol\ is 
calculated from Table 4 in Furuya et al. (2003).
The \Tbol\ ranges for the class 0, I, and II sources 
(Chen et al. 1995) are shown by the bottom bars.
\label{fig:gfevl}
}
\end{figure}

\begin{figure}
\begin{center}
\caption{Map of the intrinsic velocity width of the \HtCOp\ (1--0) emission
(greyscale and contour), obtained from the combined data.
Contour intervals are the 10\% step of
the difference between the maximum (0.71 \kms) and minimum (0.36 \kms) velocity widths.
The thick white and black contours show the 18$\sigma$ level of the total integrated intensity map of
the combined \NtwoH\ data (Figure \ref{fig:hikaku}) and
the 3$\sigma$ level of the 3\,mm continuum emission (Figure \ref{fig:contmaps}), respectively.
The green contours show the highest 3 level contours (18, 21 and 24$\sigma$ levels) 
of the total integrated intensity map of
the combined \HtCOp\ data (Figure \ref{fig:hikaku}).
The ellipses are the same as in Figure \ref{fig:tiimaps}.
\label{fig:dVcenter}
}
\end{center}
\end{figure}

\begin{figure}
\begin{center}
\caption{Upper and lower panels:~ plots of flux density in Jy vs. projected 
baseline length in $k\lambda$ for the \NtwoH\ (1--0) line data 
with the 45\,m telescope (filled circles) and the OVRO mm-array (open triangles).
The data are from \Vlsr\ $= -2.6$ \kms\ channel of the strongest
hyperfine component of $1_{23}-0_{12}$ transition.
The lower panel is a magnified plot of the upper one.
The visibilities in the plots are resampled with 0.2 and 0.4 $k\lambda$ bins
for the 45\,m and OVRO data, respectively.
\label{afig:uvamp}}
\end{center}
\end{figure}

\begin{figure}
\begin{center}
\caption{Beam patterns (upper panels) and ``dirty maps'' (lower panels) 
of the combined \NtwoH\ (1--0) data for the robustness powers of $+1, 0, -1,$ and $-2$.
The data are from the $-2.6$ \kms\ channel map of the strongest
hyperfine component of $1_{23}-0_{12}$ transition.
Contour intervals for the beams are 10\% of the peak values, starting from the $\pm 10\%$ levels. 
Dashed contours are complementary negative contours.
Contours in the dirty maps have 3$\sigma$ steps, starting from the 3$\sigma$ levels;
the 1$\sigma$ noise levels are
127.5, 82.4, 28.5, and 24.7 mJy \pbeam\ for the robustness powers of
$+1$, 0, $-1$, and $-2$, respectively.
The synthesized beam sizes of
\thetamaj\ $\times$\thetamin\ are
$10\farcs33\times 9\farcs18$,
$5\farcs81\times 5\farcs09$,
$4\farcs16\times 3\farcs57$, and
$3\farcs27\times 2\farcs66$ for the robustness powers of
$+1$, 0, $-1$, and $-2$, respectively.
The central stars in the lower panels represent the position of the 3\,mm source.
\label{afig:robust}}
\end{center}
\end{figure}

\end{document}